\documentclass[a4paper,11pt,reqno]{amsart}
\pdfoutput=1

\usepackage{amssymb,amsmath,mathabx}
\usepackage{amsfonts}
\usepackage{amssymb}
\usepackage{xcolor}
\usepackage{mathrsfs}
\usepackage{multirow}

\usepackage{tikz}
\usepackage[margin=3cm]{geometry} 
\usepackage{hyperref}
\catcode`\@11\relax
\newif\ify@autoscale \y@autoscaletrue \def\Yautoscale#1{\ifnum #1=0
  \y@autoscalefalse\else\y@autoscaletrue\fi}
\newdimen\y@b@xdim
\newdimen\y@boxdim \y@boxdim=13pt
\def\Yboxdim#1{\y@autoscalefalse\y@boxdim=#1}
\newdimen\y@linethick    \y@linethick=.3pt
\def\Ylinethick#1{\y@linethick=#1}
\newskip\y@interspace \y@interspace=0ex plus 0.3ex
\def\Yinterspace#1{\y@interspace=#1}
\newif\ify@vcenter   \y@vcenterfalse
\def\Yvcentermath#1{\ifnum #1=0 \y@vcenterfalse\else\y@vcentertrue\fi}
\newif\ify@stdtext   \y@stdtextfalse
\def\Ystdtext#1{\ifnum #1=0 \y@stdtextfalse\else\y@stdtexttrue\fi}
\newif\ify@enable@skew   \y@enable@skewfalse
\expandafter\ifx\csname enableskew\endcsname\relax
 \y@enable@skewfalse \else \y@enable@skewtrue\fi
\def\y@vr{\vrule height0.8\y@b@xdim width\y@linethick depth 0.2\y@b@xdim}
\def\y@emptybox{\y@vr\hbox to \y@b@xdim{\hfil}}
\ify@enable@skew
 \def\y@abcbox#1{\if :#1\else
   \y@vr\hbox to \y@b@xdim{\hfil#1\hfil}\fi}
 \def\y@mathabcbox#1{\if :#1\else
   \y@vr\hbox to \y@b@xdim{\hfil$#1$\hfil}\fi}
\else
 \def\y@abcbox#1{\y@vr\hbox to \y@b@xdim{\hfil#1\hfil}}
 \def\y@mathabcbox#1{\y@vr\hbox to \y@b@xdim{\hfil$#1$\hfil}}
\fi
\def\y@setdim{%
  \ify@autoscale%
   \ifvoid1\else\typeout{Package youngtab: box1 not free! Expect an
     error!}\fi%
   \setbox1=\hbox{A}\y@b@xdim=1.6\ht1 \setbox1=\hbox{}\box1%
  \else\y@b@xdim=\y@boxdim \advance\y@b@xdim by -2\y@linethick
  \fi}
\newcount\y@counter
\newif\ify@islastarg
\def\y@lastargtest#1,#2 {\if\space #2 \y@islastargtrue
  \else\y@islastargfalse\fi}
\def\y@emptyboxes#1{\y@counter=#1\loop\ifnum\y@counter>0
  \advance\y@counter by -1 \y@emptybox\repeat}
\def\y@nelineemptyboxes#1{%
  \vbox{%
    \hrule height\y@linethick%
    \hbox{\y@emptyboxes{#1}\y@vr}
    \hrule height\y@linethick}\vskip-\y@linethick}
\def\yng(#1){%
  \y@setdim%
  \hskip\y@interspace%
  \ifmmode\ify@vcenter\vcenter\fi\fi{%
  \y@lastargtest#1,
  \vbox{\offinterlineskip
    \ify@islastarg
     \y@nelineemptyboxes{#1}
    \else
     \y@ungempty(#1)
    \fi}}\hskip\y@interspace}
\def\y@ungempty(#1,#2){%
  \y@nelineemptyboxes{#1}
  \y@lastargtest#2,
  \ify@islastarg
   \y@nelineemptyboxes{#2}
  \else
   \y@ungempty(#2)
  \fi}
\def\y@nelettertest#1#2. {\if\space #2 \y@islastargtrue
  \else\y@islastargfalse\fi}
\def\y@abcboxes#1#2.{%
  \ify@stdtext\y@abcbox#1\else\y@mathabcbox#1\fi%
  \y@nelettertest #2.
  \ify@islastarg\unskip%
   \ify@stdtext\y@abcbox{#2}\else\y@mathabcbox{#2}\fi%
  \else\y@abcboxes#2.\fi}
 \newdimen\y@full@b@xdim
 \newcount\y@m@veright@cnt
\ify@enable@skew
 \def\y@get@m@veright@cnt#1#2.{%
   \if :#1 \advance\y@m@veright@cnt by 1\y@get@m@veright@cnt#2.\fi}
 \let\y@setdim@=\y@setdim
 \def\y@setdim{%
   \y@setdim@ \y@full@b@xdim=\y@b@xdim
   \advance\y@full@b@xdim by 1\y@linethick}
 \def\y@m@veright@ifskew#1{
   \y@m@veright@cnt=0 \y@get@m@veright@cnt#1.
   \moveright \y@m@veright@cnt\y@full@b@xdim}
\else
 \def\y@m@veright@ifskew#1{}
\fi
\def\y@nelineabcboxes#1{%
  \y@nelettertest #1.
  \ify@islastarg
   \y@m@veright@ifskew{#1}
    \vbox{
      \hrule height\y@linethick%
      \hbox{\ify@stdtext\y@abcbox#1\else\y@mathabcbox#1\fi\y@vr}
      \hrule height\y@linethick}\vskip-\y@linethick
  \else
   \y@m@veright@ifskew{#1}
    \vbox{
      \hrule height\y@linethick%
      \hbox{\y@abcboxes #1.\y@vr}%
      \hrule height\y@linethick}\vskip-\y@linethick
  \fi}
\def\young(#1){%
  \y@setdim%
  \hskip\y@interspace%
  \y@lastargtest#1,
  \ifmmode\ify@vcenter\vcenter\fi\fi{%
  \vbox{\offinterlineskip
    \ify@islastarg\y@nelineabcboxes{#1}%
    \else\y@ungabc(#1)%
    \fi}}\hskip\y@interspace}
\def\y@ungabc(#1,#2){%
  \y@nelineabcboxes{#1}%
  \y@lastargtest#2,
  \ify@islastarg\y@nelineabcboxes{#2}%
  \else\y@ungabc(#2)%
  \fi}
\catcode`\@12\relax
 
\theoremstyle{definition}

\newcommand{\BE}{\begin{equation} \begin{array}{c}}
\newcommand{\EE}{\end{array}\end{equation}}

\newcommand{\LL}{\mathscr L}

\def\Phibar{ \overline{\Phi} }
\def\Phibarbar{ \overline{\overline{\Phi}} }

\def\one{\hbox{{1}\kern-.25em\hbox{l}}}

\def\nn{\nonumber}

\def\uot{ \underline{\otimes} }

\def\uA{ \underline{A} }

\def\uE{ \underline{E} }
\def\uB{ \underline{B} }

\def\uS{ \underline{S} }
\def\uR{ \underline{R} }
\def\uG{ \underline{G} }
\def\Gb{\overline{G}}
\def\oG{\overline{G}}
\def\oA{\overline{A}}
\def\oZ{\overline{Z}}
\def\uGb{ \overline{\underline{G}} }
\def\uSb{ \overline{\underline{S}} }
\def\uRb{ \overline{\underline{R}} }
\def\uoE{ \underline{\overline{E}} }
\def\uoB{ \underline{\overline{B}} }
\def\uoA{ \underline{\overline{A}} }
\def\uoR{ \underline{\overline{R}} }
\def\uoS{ \underline{\overline{S}} }

\def\uoG{ \underline{\overline{G}} }

\def\uF{ \overline{F} }
\def\ulx{ \underline{x} }
\def\ul{ \underline{\ell} }
\def\uell{ \underline{\ell} }
\def\unab{ \underline{\nabla} }
\def\ri{\mathrm{i}}
\def\th{\textstyle{\frac 12}}
\def\ta{\texttt{a}}
\def\tb{\texttt{b}}
\def\tta{\texttt{a}}
\def\ttb{\texttt{b}}

\def\ttd{\texttt{d}}
\def\ttD{\texttt{D}}

\definecolor{vertclair}{RGB}{0,153 0}

\definecolor{gris}{RGB}{153,153,153}

\def\yng#1{\scalebox{.5}{\raisebox{-0.1cm}{$\young(#1)$}}}

\title[Antisymmetric tensor fields] 
 {  Antisymmetric tensor fields: actions, symmetries\\ and \\first order Duffin-Kemmer-Petiau formulations
}

\author[Jarvis and Thierry-Mieg]{Peter D. Jarvis${}^1$ and Jean Thierry-Mieg${}^2$}
\address{${}^1$School of Natural Sciences (Mathematics and Physics), University of Tasmania, PO Bag 37, Hobart, Tasmania 7001, Australia
; 
\mbox{${}^2$NCBI}, National Library of Medicine, National Institute of Health, 8600 Rockville Pike, Bethesda MD20894, U.S.A.
}
\email{${}^1$peter.jarvis@utas.edu.au (Alexander von Humboldt Fellow), ${}^2$mieg@ncbi.nlm.nih.gov}

\date{\today }
\begin{document}

\begin{abstract}
  Analyzing the representations of the Lorentz group, we give a systematic count and construction of all the possible Lagrangians describing an antisymmetric rank two tensor field.
  The count yields two scalars: the gauge invariant Kalb-Ramond model, equivalent to the sigma model and familiar from super gravity and string theory, and the conformally invariant
  Avdeev-Chizhov model, which describes self-dual tensors.
The count also includes a third invariant, a pseudoscalar, which is an antisymmetrized form of the Avdeev-Chizhov Lagrangian, first noticed in the $SU(2/1)$ superalgebraic model of the weak interactions. This model is also conformally invariant, and naturally implements the Landau $CP$ symmetry.
  Then, by extending the Duffin-Kemmer-Petiau 10 component formalism, we recover the model Lagrangians as first order systems.
  To complete the analysis we classify all local Lorentz invariant potentials (mass terms and quartic couplings)
  for charged antisymmetric tensor fields coupled to a Yang-Mills field.
\end{abstract}

%\date{\LARGE \textbf{DRAFT}}
\normalsize
\maketitle
\vfill
\tableofcontents
%\vfill

%\footnotetext{${}^*$ Alexander von Humboldt Fellow}
%\tableofcontents
%\pagebreak

\section{Introduction and main results.} 
\label{sec:Intro}
\mbox{}

\noindent
The theoretical project of constructing invariant actions for `higher spin' relativistic fields can plausibly be claimed to have originated with the Klein-Kaluza models, in curved space, but was subsequently shaped by Dirac's formulation for spin-$\textstyle{\frac 12}$ and generalizations \cite{Dirac1936relativistic}. Famous early examples include the Proca spin-1 equation
\cite{proca1936theorie}, and, for example the treatment by Fierz and Pauli of spin-$\textstyle{\frac 32}$ and spin-2\,, \cite{fierz1939relativistic}, but the topic has spawned an enormous and still expanding literature.
The successful validation of the elementary particle standard model
has led to the quantum field-theoretical focus on ``fundamental'' entities ,
as currently understood to be quarks and leptons (spin-$\textstyle{\frac 12}$ Fermions), together with vector exchange particles (spin-$\textstyle{1}$ gauge Bosons), being augmented by an additional, apparently fundamental, spin-$\textstyle{0}$ entity, the Higgs Boson, with deep implications for the origin of mass, and cosmology.
 
While ever the full experimental verification of the physics of the Bosonic sector and symmetry breaking phenomena is incomplete, it is arguably prudent on a theoretical level to reexamine possible alternative presentations of the corresponding physical models. It is this question that we take up in the present paper.

In this vein we reconsider the case of local relativistic antisymmetric 
tensor fields (rank 2 tensors in 4 dimensional Minkowski space). In the guise of  higher spin  gauge models \cite{ogievetsky1967notoph}, such
degrees of freedom are a known concomitant of various string theories and dualities
(see for example \cite{KalbRamond1974}), and indeed have been much studied 
in topological quantum field theory, in applications to geometric invariant theory \cite{blau1991topological}. Skew symmetric tensor gauge fields have also been invoked in connection with the QCD $U(1)$ problem \cite{Hata1981skew}.
Along with the graviton and certain scalar multiplets, the two-form potentials also constitute the basic gauge multiplet of so-called double field theory (see for example \cite{Choi_2015}). Moreover, a different formulation,
of conformally invariant antisymmetric tensor fields, proposed and developed some time ago \cite{AvdeevChizhov1994} has also been explored
as a possible accompaniment of phenomenological scalar degrees of freedom in the standard model symmetry breaking Higgs sector \cite{Chizhov2011}\, 
(for technical aspects see also  \cite{LemesRenanSorella1995II,wetterich2008quantization}). 

In section \ref{sec:Lorentz_kinetic}  below we present a systematic count of all local invariant combinations (one dimensional representations) which are quadratic in a generic antisymmetric tensor and in derivatives, thereby enumerating all admissible linearly independent kinetic terms for the free fields. The tensor analysis is based 
on the analysis of representations of the special Lorentz group $SO(3,1)$\,, but also on group character theory in the full Lorentz group $O(3,1)$\,. The result of this counting 
is that in symmetric coupling, as to be expected, there are just two linearly independent Lorentz scalar candidate invariants, which reproduce (in appropriate linear combinations) the known gauge invariant two-form potential,
and conformally invariant antisymmetric tensor models referred to above. Furthermore, the count of one dimensional representations includes two additional Lorentz pseudo-scalars:
one symmetrically coupled, and one antisymmetrically coupled. The antisymmetrically coupled pseudo-scalar, which is also conformally invariant, has also been 
exploited in our recent work
\cite{Thierry_Mieg_Jarvis_2021}\,, in which
``chiral Bosons'' (c.f. \cite{wetterich2008quantization}) appear as part of the
$SU(2/1)$ superalgebraic extension of the electroweak gauge sector. (See section \ref{sec:Conclusions} for further discussion)\,. 

In section \ref{sec:Lagrangian_section}, we flesh out the previous classification by giving free Lagrangians
with the desired properties. From the literature, these include the
two symmetric scalar Lagrangians: the Kalb-Ramond gauge invariant model $\LL^{KR}$ \cite{KalbRamond1974}  and the Avdeev-Chizhov self-dual tensor model $\LL^{AC}$ \cite{AvdeevChizhov1994},
and thirdly the antisymmetric self-dual pseudo-scalar Lagrangian $\LL^{CP}$ from \cite{Thierry_Mieg_Jarvis_2021}.
The remaining, fourth case is a trivial, 
symmetric pseudo-scalar Lagrangian, describing a collection of 6 independent fields.

Amongst the contributions to the early work on relativistic wave equations should also be included 
the papers of Duffin, Kemmer and Petiau (DKP) \cite{duffin1938characteristic,
Kemmer1939,petiau1938university}\,.
In particular, the seminal paper of Kemmer presented a general first order equation,  involving the Kemmer $\beta$ matrices, a weaker algebraic structure than that of the Dirac matrices.
Two different cases, with 5- and 10-component wave functions,
provided (at least for free fields) for formulations equivalent to 
the Klein-Gordon (massive complex scalar) spin-$\textstyle{0}$ equation and the Proca
(massive complex vector) spin-$\textstyle{1}$ equation. 
%
%In the paper [cite-Kemmer], Kemmer also speculated that in the 10-component wave function, which in its standard formulation consisted of a 4 component vector part and its 6 component curl, the roles could be inverted, to encompass instead a 6 component antisymmetric tensor, together with its totally antisymmetric curl (as an axial vector).

%
%The realization of Kemmer's suggestion, and thus potentially opening a new arena for describing the physics of the antisymmetric tensor fieldsis the topic of the second part of the present paper. 
In section \ref{sec:DKP} and appendix \ref{sec:KemmerAlgebra} below, we present an extension of the 10 component Duffin-Kemmer-Petiau system.
Specifically, we show that there are two `twisted' variants of the
Duffin-Kemmer-Petiau local action
providing first order equations, which after elimination of auxiliary fields, precisely reproduce the two known physical models identified above in the count for the symmetric coupling (the two-form gauge potential and the conformally invariant antisymmetric tensor field, respectively). For antisymmetric field coupling, we demonstrate a third variant, which recovers the pseudo-scalar term identified above. 
As an illustration of the method, the standard Proca system for a massive vector field is given in first order DKP form in the appendix \ref{subsec:DKP_Proca}. The corresponding details for the two form gauge potential and the two antisymmetric tensor field cases are given in appendix \ref{subsec:DKP_2FGAST}. The conformal invariance of these two models is discussed in appendix \ref{sec:ConfDKPproof}. 

Moving from kinetic terms to couplings, for completeness we also provide in section  \ref{sec:Lorentz_potentials} a count of Lorentz invariant local interactions, (``mass'' terms and quartic potentials), for physically relevant cases
of a complex scalar and complex antisymmetric tensor, invariant under global internal $U(n)\cong SU(n)\times U(1)$\,, in the fundamental $n$-dimensional representation.

In section \ref{sec:Conclusions}, we summarize our main results and discuss the implications of these models for physics.

\section{Lorentz invariant tensor polynomials: kinetic energy}
\label{sec:Lorentz_kinetic}
%\mbox{}\\
%\begin{quotation}
%
%\noindent
%Summary of character methods: \\
%use notation $(j_1,j_2)$ for $SO(3,1) \cong sl(2)\times sl(2)$ irreps and ${}_S, {}_A$ for symmetrization and  antisymmetrization\,. 
%Antisymmetric tensor $Z_{\mu\nu} := Z_++Z_- \cong (1,0)+(0,1)$\,.
%Gradient: $\partial_\mu := K \cong (\textstyle{\frac 12}, \textstyle{\frac 12})$\,;
%tensor gradient $K.T$%\\%[.2cm]
%\mbox{}\\
%\noindent
%\emph{Claims:} Results as in the intro.
%%\emph{Claim 1:} Symmetric kinetic term $(K.T)_S \cong K_A.T_A + K_S.T_S$ contains 5 Lorentz invariants,
%%deriving from $K_S.(T_\pm)_S$ (two from each) and one from $K_S.(T_+.T_-)$.\\[.2cm]
%%\emph{Claim 2:} Antisymmetric kinetic term $(K.T)_A \cong K_A.T_S + K_S.T_A$ contains one
%%Lorentz invariant (from $K_S.(T_+.T_-)$).\\[.2cm]
%%\emph{Claim 3:} the new Lorentz invariant (Claims 1 \& 2) is a pseudoscalar under the full Lorentz group $O(3,1)$\,.
%\\
%\mbox{}\hfill $\Box$
%\end{quotation}
Relativistic fields in Minkowski space are labeled as
representations of the Lorentz group by their associated spins. For the special Lorentz group $SO(3,1)$\,, we exploit the isomorphism
%\footnote{Finite dimensional representations can be handled
%at the level of the four dimensional complex groups irrespective of real forms.} 
with the direct product $sl(2)\times sl(2)$\,, and denote irreducible representations
by pairs $(j_1,j_2)$ with half-integer spins $j_1, j_2 = 0, \textstyle{\frac 12}, 1, \cdots$\, and dimension $(2j_1+1)(2j_2+1)$\,. Thus a Lorentz vector
is the representation $(\textstyle{\frac 12}, \textstyle{\frac 12})$\,, while 
 $(1, 1)$ represents a traceless symmetric tensor (such as the gravitational metric in some gauges)\,. Finally, an antisymmetric tensor, of dimension 6, is a reducible
representation $(1,0)+(0,1)$\, (see below). 
%In the sequel (in view of the multiplicity of intermediate tensor types in the counting) we do not use the language of forms, but classify
%actions by counting local relativistic invariant monomials as integrands for the standard
%four dimensional Riemannian measure.
%\pdj{MOVE TO CONCL BUT KEEP $d^4x$ COMMENT}

%In explicit constructions [citation: Schur (software), King (guts paper); Wybourne-Gidding; Hehl], 
For enumerating candidate kinetic terms we take local invariants quadratic in derivatives and fields, of the form
\begin{equation}
\label{eq:CombCount}
\mathscr{L} = \texttt{C}^{\mu\nu\alpha\beta\gamma\delta} \partial_\mu
T_{\alpha \beta}\partial_\nu T_{\gamma\delta}\, + \mathrm{h.c.},
\end{equation}
where\footnote{We use standard notation for Minkowski 
space with coordinates $x^\mu =(x^0, x^i) = (ct, \ulx)$\,, metric
$\eta_{\mu\nu}=\mathrm{diag}(1,-1,-1,-1)$ and Levi-Civita
tensor $1=\varepsilon_{0123}=-\varepsilon^{0123}$\,.} $T_{\mu\nu}=-T_{\nu\mu}$\, and $\texttt{C}^{\mu\nu\alpha\beta\gamma\delta}$ is any numerical tensor made from 
(monomials in) the invariant tensors
%\footnote{If there is no dependence on 
%$\eta_{\kappa \lambda}$ or in general on a background metric, then the model is ``topological''.} 
$\eta_{\kappa \lambda}$ and 
$\varepsilon_{\rho \sigma \tau \upsilon}$\,. 
%Up to integration by parts, it is immaterial for counting purposes where the derivatives act; for expediency in comparison with standard forms we assume that the gradients each act on only one of the two $Z$ terms.
With this in mind we have the derivative terms contributing a symmetric tensor
(of dimension 10)
%\footnote{Working in 
%doubled coordinates such as complex Minkowski space, or extended phase space
%$x \pm ip$, would allow more general symmetries, for example via non-holomorphic
%terms, or antisymmetrized iterated Hirota derivatives.} 
with the two $T$ terms
providing additional factors in the overall tensor product, written symbolically as a product of irreducible representation labels (or group characters),
\begin{equation}
\label{eq:KineticCharacter}
\big( (0,0)+(1,1)\big)\!\cdot\! \big((1,0)+(0,1)\big)^2\,.
\end{equation}
Thus far we have suppressed (global, or eventually local) internal symmetry transformations, which would
require appending to the fields an additional internal suffix ${}^{\ta,\tb},\cdots,$\,.
Here we keep this implicit, but to accommodate later extensions (see section
\ref{sec:Lagrangian_section} and \ref{sec:Lorentz_potentials} below) we separately take account of which invariants arise from symmetric, or antisymmetric, couplings (corresponding to orthogonal or symplectic representations of internal symmetry groups, respectively).\footnote{The role of the Hodge
dual ${}^*T$ in the projection of a tensor $T$ into self-dual and anti self-dual parts $(1,0)$ and $(0,1)$ is explained below in relation to explicit local forms of the invariants. } 

A simple count  (appendix \ref{subsec:kinetic}) shows that there are four invariants in total: three symmetric, and one antisymmetric.

We now augment the analysis, by identifying the above invariants under the full
Lorentz group $O(3,1)$\,. 
At this level the machinery of tensor representations is somewhat more intricate than for $SO(3,1)$\,, and relevant details are provided in appendix \ref{subsec:kinetic} below, based on formal group character manipulation. 

Table \ref{tab:SO31O31Count}\, shows how the $SO(3,1)$ count is refined
under $O(3,1)$\,. Clearly, under restriction the latter count (of one dimensional representations) must reduce to that for the special Lorentz group. In symmetric coupling we find two scalars and one pseudo-scalar,
while the single $SO(3,1)$ invariant in antisymmetric coupling is a Lorentz pseudo-scalar.

\begin{table}[tbp]
\begin{tabular}{|c|c|c|c|c|}
\hline 
%& \multicolumn{2}{|c|}{ } & \multicolumn{2}{|c|}{ } \\ \hline
& \multicolumn{2}{|c|}{symmetric} & \multicolumn{2}{|c|}{antisymmetric} \\ \hline 
$SO(3,1)$ & \multicolumn{2}{|c|}{3} & \multicolumn{2}{|c|}{1} \\ \hline
\multirow{2}{*}{$O(3,1)$}& scalar & pseudo-scalar & scalar & pseudo-scalar \\
\cline{2-5} & 2 &1&0&1 \\
\hline
\end{tabular}
\mbox{}
\caption{
Count of symmetric and antisymmetrically coupled $SO(3,1)$ and $O(3,1)$ invariant kinetic terms.}
\label{tab:SO31O31Count}
\end{table}

Capitalizing on previous knowledge, we present below for each of these cases, a corresponding Lagrangian, written in the standard covariant way\,.
In this context, the interest of the present section is the proof that in four dimensions our enumeration of these Lagrangians is complete.

\section{Lorentz invariant tensor polynomials: free Lagrangians.}
\label{sec:Lagrangian_section}
In the previous section we established that in four dimensional Minkowski space there are precisely four linearly independent local terms which are suitable ingredients for kinetic terms of antisymmetric tensor Lagrangians. 

It is straightforward to construct, in standard relativistic tensor notation 
(compare (\ref{eq:CombCount}))\, an explicit basis of local monomials which, given the count, must provide the ingredients for any physical models. Four such candidates, $\mathscr{L}^\Box$\,, $\mathscr{L}^\nabla$\,, ${}^*\mathscr{L}^\Box$\,,and ${}^*\mathscr{L}^\nabla$ are listed in table \ref{tab:candidates}, with properties in accord with table \ref{tab:SO31O31Count} above.
\begin{table}[tbp]
\begin{tabular}{|c|c|c|c|}
\hline
&&&\\
&$\mathscr{L}^\Box$ & $T_{\mu\nu}\partial^\rho\partial_\rho T^{\mu\nu}$& $(T\Box T)$\\
&&&\\
\hline
&&&\\
&$\mathscr{L}^\nabla$&
$(\partial^\rho T_{\rho \mu})\eta^{\mu\nu}(\partial^\sigma T_{\sigma \nu})$
&$(\partial\!\cdot\! T)(\partial\!\cdot\!{}T)$\\
&&&\\
\hline
&&&\\
$P$&${}^*\mathscr{L}^\Box$ &${}^*T_{\mu\nu}\partial^\rho\partial_\rho T^{\mu\nu}$& $({}^*T\Box T)$\\
&&&\\
\hline
\hline
&&&\\
$P$&${}^*\mathscr{L}^\nabla$&
$(\partial^\rho {}^*T_{\rho \mu})\eta^{\mu\nu}(\partial^\sigma T'_{\sigma \nu})$&
$(\partial\!\cdot\!{}^*T)(\partial\!\cdot\!{}T')$\\
&&&\\
\hline
\end{tabular}
\caption{Explicit local candidates $\mathscr{L}^\Box$\,, $\mathscr{L}^\nabla$\,, ${}^*\mathscr{L}^\Box$\,, and ${}^*\mathscr{L}^\nabla$ for  $SO(3,1)$ and $O(3,1)$ invariant kinetic terms. The notation ${}^*T$ denotes the Hodge dual (see text). The first three rows are for symmetric coupling,
and the last row (vanishing if $T'\!=\!T$) is for antisymmetric coupling.  For discussion purposes, an informal representation of each candidate term is also listed.   }
\label{tab:candidates}
\end{table}
\mbox{}

We now review and discuss various Lagrangians incorporating these terms, and which (as mentioned in the introduction) which have been considered as models for the physics of antisymmetric tensor fields.
Note that the third invariant in table \ref{tab:candidates}, the pseudoscalar ${}^*\mathscr{L}^\Box = {}^*T\Box T$\,, does not appear to have been used in model building, and will not be discussed further. The symmetrically coupled, scalar, contributions are incorporated in the Kalb-Ramond and Avdeev-Chizhov models 
\cite{KalbRamond1974}, \cite{ogievetsky1967notoph}, \cite{AvdeevChizhov1994},
\begin{align}
\label{eq:KR_AC_real}
{\mathscr L}^{KR}=&\, -\mathscr{L}^\Box - 2\mathscr{L}^\nabla= 
(\partial^\rho T_{\mu\nu})(\partial_\rho T^{\mu\nu}) -
2(\partial^\rho T_{\rho \mu})\eta^{\mu\nu}(\partial^\sigma T_{\sigma \nu})\,;\\
{\mathscr L}^{AC}=&\, -\mathscr{L}^\Box - 4\mathscr{L}^\nabla\,= 
(\partial^\rho T_{\mu\nu})(\partial_\rho T^{\mu\nu}) -
4(\partial^\rho T_{\rho \mu})\eta^{\mu\nu}(\partial^\sigma T_{\sigma \nu})\,.
\end{align}
Note that ${\mathscr L}^{AC}$ can be re-written \cite{AvdeevChizhov1994} in terms of the self-dual and anti self-dual projections of a complex antisymmetric tensor $Z_{\mu\nu}$\,,
\[
Z := T + \mathrm{i} {}^*T\,,\qquad \oZ := T - \mathrm{i}{}^*T\,, 
\]
where ${}^*T_{\mu\nu} := \textstyle{\frac 12}\varepsilon_{\mu \nu \rho \sigma}T^{\rho \sigma}$ is the Hodge dual such that ${}^{**}T = -T$ \, in Minkowski space, with ${}^*Z = -\mathrm{i} Z$\,, ${}^*\oZ = +\mathrm{i}\oZ$\,. With these definitions we find 
\cite{AvdeevChizhov1994}
\begin{align}
\label{eq:AC_complex}
{\mathscr L}^{AC} = &\,
\partial^\rho \oZ_{\rho \mu} \eta^{\mu\nu}\partial^\sigma Z_{\sigma \nu}\,.
\end{align}
The Kalb-Ramond model does not admit an analogous rearrangement, but instead
can be re-written
\begin{align}
\label{eq:KR_delta_det}
{\mathscr L}^{KR}=  \textstyle{\frac16}\delta^{\lambda \mu \nu}{}_{\rho \sigma\tau} 
\partial_{\lambda}T_{\mu\nu}\partial^\rho T^{\sigma\tau}
\end{align}
in terms of the totally antisymmetric determinant 
invariant\footnote{We have 
$\delta^{ \lambda\mu\nu}{}_{\rho \sigma\tau} =
\varepsilon^{\kappa\lambda\mu\nu}
\varepsilon_{\kappa\rho \sigma\tau} $\,; recall from (\ref{eq:CombCount}) that scalar and pseudoscalar Lagrangian
densities must be even or odd in $\varepsilon_{\alpha\beta\gamma\delta}$\,, respectively.\,}
$
\delta^{ \lambda\mu\nu}{}_{\rho \sigma\tau} \!=\! 
\delta^\lambda{}_{\rho} \delta^{\mu}{}_{\sigma}  \delta^{\nu}{}_{\tau} \! \pm\!\cdots %\mbox{\textrm{perms}}\Big)\,.
$\,.

Rather than expanding ${}^*\mathscr{L}^\nabla=(\partial\!\cdot\! \oZ)(\partial\!\cdot\!  Z)$ in terms of projections as above, consider now the corresponding expansion starting with two distinct fields $Z, Z'$ and projections $T$\,, $T'$ and ${}^*T$\,, ${}^*T'$\,. This reproduces the analogue of ${\mathscr L}^{AC}$\, (symmetrized in $T,T'$)\,, but also returns a cross term, which is the antisymmetric pseudoscalar listed in table \ref{tab:candidates} above,
\begin{equation}
\label{eq:StarLdiv}
{}^*\mathscr{L}^\nabla :=\mathrm{i}
(\partial^\rho {}^*T_{\rho \mu})\eta^{\mu\nu}(\partial^\sigma T'_{\sigma \nu})\,.
\end{equation}
Endowed with a multiplet of complex fields $Z^{\tta}$\,, $a=1,2,\cdots$\,, 
belonging to a representation of some internal symmetry group, with antisymmetric 
quadratic invariant $\kappa_{\tta\ttb}=-\kappa_{\ttb\tta}$\,, this becomes the model 
of antisymmetric tensors
first identified in \cite{Thierry_Mieg_2020,Thierry_Mieg_2021,Thierry_Mieg_Jarvis_2021} via Fermion loop corrections in a superalgebra-enhanced standard model, augmented by antisymmetric tensor fields in the gauge sector\footnote{The extended (non-Abelian) Avdeev-Chizhov model is of this form, but with
a symmetric quadratic invariant $\kappa_{\tta\ttb}=+\kappa_{\ttb\tta}$\,.}:
\begin{align}
\label{eq:CP_complex}
\mathscr{L}^{CP} :=
\mathrm{i}(\partial^\rho {}^*T^\tta_{\rho \mu})\eta^{\mu\nu}(\partial^\sigma T^\ttb_{\sigma \nu})\kappa_{\tta\ttb}\,.
\end{align}

Importantly, the models ${\mathscr L}^{KR}$, ${\mathscr L}^{AC}$ and ${\mathscr L}^{CP}$ which we have identified are distinguished by their symmetries. Evidently, ${\mathscr L}^{KR}$ is gauge invariant\footnote{In the language of exterior forms, the action corresponding to the Lagrangian density ${\mathscr L}^{KR}$ is proportional to $\int {}^*dT \wedge dT$\,, which is manifestly gauge invariant under $T\rightarrow T + dX$\,.}, while ${\mathscr L}^{AC}$ (containing a 
different admixture of ${\mathscr L}^{\Box}$ and ${\mathscr L}^{\nabla}$) is not. Vice versa, as noted in the
original literature, ${\mathscr L}^{AC}$ is invariant under the conformal group acting in four-dimensional Minkowski space, whereas ${\mathscr L}^{KR}$ is not. 

However, ${\mathscr L}^{KR}$, ${\mathscr L}^{AC}$
(and also ${\mathscr L}^{CP}$) are specific instances of hierarchies of models of generalized antisymmetric tensors, of various ranks $p$\,, and space-time dimension $D$. For example,
${\mathscr L}^{KR}$ itself is a generalization to rank 2 of Maxwell electromagnetism, which is well known to be conformally invariant \cite{bateman1910transformation} as well as being gauge invariant; however for antisymmetric rank 2, conformal invariance holds in $D=6$ rather than $D=4$. For a comprehensive
discussion we refer the reader to \cite{ThierryMiegJarvis2023conformal}, where it is also proven that the generalization of 
${\mathscr L}^{CP}$\,, for self-dual fields, is conformally invariant in any dimension $D=2p$ in Minkowski space\,. Further details are beyond the scope of this paper; however, given the significance of the results, in appendix \ref{sec:ConfDKPproof} we provide for completeness, a derivation of conformal invariance of ${\mathscr L}^{CP}$ and 
${\mathscr L}^{AC}$ within the four dimensional first order formulation. 

\section{10 component DKP formulation for antisymmetric tensor fields}
\label{sec:DKP}
It is well known that 
the wave equations associated with the Dirac algebra
\[
\{\gamma_\mu,\gamma_\nu\}=2\eta_{\mu \nu}\,,
\]
together with those for the Duffin-Kemmer algebra
\[
\beta_\mu \beta_\rho \beta_\nu + 
\beta_\nu \beta_\rho \beta_\mu = \eta_{\mu \rho}\beta_\nu +\eta_{\nu \rho}\beta_\mu \,,
\]
provide the only single-mass instances of the general class of $SO(5)$-related Bhabha wave equations. The Dirac equation of course pertains to Fermionic fields of spin-$\textstyle{\frac 12}$\,, while (as reviewed above), the DKP equation either to Bosonic spin 0 (for 5 component wave functions) or spin 1\, (for 10 component wave-functions, in the standard picture)\footnote{These
cases correspond to the irreducible 5- and 10- dimensional representations of the
Kemmer algebra and, together with the trivial 1-dimensional representation, saturate the dimension of the adjoint representation $126=1^2 + 5^2 + 10^2$ in accordance with Wedderburn's theorem.}\,. Both cases have been much studied as alternatives to
the complex scalar Klein-Gordon model and the 
complex vector Proca model, both in phenomenological applications, and also in relation to their equivalence to these standard systems, especially in the interacting case
or in curved backgrounds (for an overview see \cite{KrajcikNieto1977historical}\,,
 for classical solutions see for example \cite{NedjadiBarrett1993,NedjadiBarrett1994}\,.
A detailed analysis has been presented in \cite{Kruglov2010kalb,kruglov2011field},  
and for investigation at the second quantized level see 
\cite{BeltranPimentelSoto2020}\,).

An explicit representation of the Kemmer $\beta$-matrices in the $10\times 10$ case
is conveniently written in $3 \!+\!3 \!+\!3\!+\!1$ block form as follows\footnote{Here $e_i$\,, $i=1,2,3$ are standard 3-component unit column vectors, $f_i = \smash{{}^\top e_i}$ the corresponding row vectors, and $\ell_i :=\textstyle{\frac 12}\varepsilon_{ijk} \ell_{jk}$ elementary $3\times 3$ rotation generators. (For details see appendix \ref{sec:KemmerAlgebra}\,).} :
%For completeness, we give explicit representations of the Kemmer $\beta$-matrices for (in $1\!+\!3\!+\!1$ block form) the $5\times 5$\,: 
%\begin{align*}
%\beta^0=&\,\left[ \begin{array}{c|c|c} 
%\cdot & \cdot & 1\\ \hline
%\cdot & \cdot & \cdot \\ \hline
%1 & \cdot & \cdot \\
%\end{array}\right]\,, \quad
%\beta^i=\left[ \begin{array}{c|c|c} 
%\cdot & \cdot & \cdot \\ \hline
%\cdot & \cdot & e_i  \\ \hline
%\cdot & -f_i & \cdot \\
%\end{array}\right]\,;
%\end{align*}
%and (in $3_1\!+3_2\!+\!3_3\!+\!1$ block form) $10\times 10$\,: 
\begin{align}
\label{eq:Beta10Defs}
\beta^0 \!=\! \left[ \begin{array}{ccc|c} 
\cdot & \cdot & 1 & \cdot  \\
\cdot & \cdot & \cdot & \cdot \\
1 & \cdot &   \cdot& \cdot \\ \hline
\cdot &\cdot &\cdot &\cdot 
\end{array}\right]\,,\quad
%%
%\beta^i \!=\!&\, \left[ \begin{array}{ccc|c} 
%\cdot & \cdot & \cdot & \beta e_i\\
%\cdot & \cdot & {{\alpha' }}\ell_i & \cdot \\
%\cdot & \alpha\ell_i &\cdot & \cdot\\ \hline
%\beta' f_i &\cdot &\cdot &\cdot 
%\end{array}\right]\,;\quad
%
\beta^i \!=\!&\, \left[ \begin{array}{ccc|c} 
\cdot & \cdot & \cdot &  e_i\\
\cdot & \cdot & \ell_i & \cdot \\
\cdot & \ell_i &\cdot & \cdot\\ \hline
 -f_i &\cdot &\cdot &\cdot 
\end{array}\right]\,.\quad
\end{align}
The three triplet plus singlet partitioning reflects the reducible Lorentz structure of the DKP wave-function at the level of the rotation group: namely, antisymmetric tensor
(dimension 6) plus 4-vector.
%\footnote{The corresponding $5\times5$ representation in  $1\!+\!3\!+\!1$ block form, representing 4-vector plus singlet, is provided for completeness in appendix \ref{sec:KemmerAlgebra}\,.}.
%This is given in a basis-independent way by analyzing the spectrum of the Lorentz group Casimir operators\footnote{With eigenvalues $j_1(j_1+1)\pm j_2(j_2+1)$\,, respectively.} $C_2 := J^{\mu\nu}J_{\mu\nu}$ and 
%$C_2':= \varepsilon^{\mu \nu\rho\sigma}J_{\mu\nu}J_{\rho\sigma}$
%quadratic in the $SO(3,1)$ generators $J_{\mu\nu} := \textstyle{\frac 14}{[}\beta_\mu,\beta_\nu{]}$\,. 
%A further ingredient necessary for constucting and interpreting Lorentz invariant local actions in the DKP formalism is the identification of t
The parity transformation, such that for the 10-component wave-function $\Phi'(ct, -\ulx) = -\eta\Phi(ct, \ulx) $\,, leads to the standard parity matrix \cite{Kemmer1939}
\begin{align}
\label{eq:EtaDef}
\eta:= &\,2(\beta^0)^2-1 = \mathrm{diag}(1,-1,1,-1)\,
\end{align}
and in the usual way to the conjugate wave-function
$\Phibar := \Phi^\dagger \eta$\,. 

%As presaged in the introduction, one aim of this section is to realize the old observation of Kemmer [cite-Kemmer] that the standard Proca spin-1 interpretation of the 10-component DKP equation can be inverted, to become a theory of an antisymmetric tensor field, with the 4 component vector its
%antisymmetric curl (as an axial vector). In fact, as we shall show presently, different variants are possible, which will turn out both to recover known models of antisymmetric tensors, and also an additional possible model. In all, we confirm 
%that these cases indeed exhaust the naive count of admissible Lorentz-invariant actions from section \ref{sec:Lorentz_kinetic} above. 

In appendix \ref{subsec:DKP_Proca} we demonstrate for completeness the well-known standard DKP derivation of the Proca action for a (free) massive complex spin-1 field\,. 
%(the 
%corresponding analysis for the derivation of the complex scalar massive Klein-Giordon equation in the 5-component case is also given). 
%Before elaborating on the generalized DKP formalism, it is useful to note some features of the standard case.  
%Recalling that we require three 3-vectors and a singlet, a suitable notation
%for the 10 component wavefunction is
%\begin{align}
%\label{eq:MaxwellLabels}
%\Phi \cong {}^\top\big(\uE, \uB, \uA,A_0\big)\,.
%\end{align}
%With standard parity assignment as in (\ref{eq:EtaDef}), this is suggestive of 
%the subsequent identifcation of $\uE, \uB$ as the vector and pseudovector electric and magnetic parts of the Maxwell field strength tensor and is indeed consistent.
%Note that the opposite parity choice $-\eta$ would also be consistent, with the 
%electric and magnetic parts acquiring instead the opposite parities as three-vectors, so that the theory is that of a massive complex spin-1 axial vector meson.
%For the generalized DKP system, a crucial ingredient will be the alternative assignment
%(for derivation see below)
%\begin{align}
%\label{eq:EtaPrDef}
%\eta':=  \mathrm{diag}(-1,1,1,-1)\,
%\end{align}
%whereby the final 4-vector becomes an axial vector (as required if it is to be identified with the curl of
%an antisymmetric tensor), but without any change in the parity assigments to the analogues of the first two three-vectors.
Kemmer \cite{Kemmer1939} pointed out that the roles of the antisymmetric tensor (field strength) and vector in the 10 component DKP wave-function could be reversed
(now with an axial vector as field strength tensor). 
In this vein we proceed to develop a corresponding extended DKP formalism as follows. 
Firstly note that the pseudo-scalar object
$\beta_5 := \textstyle{\frac 18}\varepsilon^{\mu \nu\rho\sigma}
{[}\beta_\mu,\beta_\nu{]}{[}\beta_\rho,\beta_\sigma{]} \equiv
\textstyle{\frac 12}\varepsilon^{\mu \nu\rho\sigma}
\beta_\mu\beta_\nu\beta_\rho\beta_\sigma$ is the analogue for the 
Kemmer algebra of the pseudo-scalar $\gamma_5 = \gamma_0\gamma_1\gamma_2\gamma_3$ in the Dirac case, and takes the form
\footnote{$\beta_5 \equiv 0$ in the 5 dimensional representation.}
\begin{align}
\label{eq:Beta5Def}
{\beta_5} \!=\! &\,\left[ \begin{array}{ccc|c} 
\cdot & 1 & \cdot & \cdot  \\
-1& \cdot & \cdot & \cdot \\
\cdot & \cdot &   \cdot& \cdot \\ \hline
\cdot &\cdot &\cdot &\cdot 
\end{array}\right]\,.
\end{align}
We now assume that in an extension of the DKP equation,
the Kemmer matrix $\beta^\mu$ may be replaced by an analogue taken from 
the enveloping algebra. 
Here we investigate the choice
\begin{align}
\label{eq:Beta5check}
\widecheck{\beta}{}^\mu := &\,{[}\beta_5, \beta^\mu{]}\,,
\end{align}
from which
\begin{align}
\label{eq:NewBetaList}
\widecheck{\beta}^0 \!=\! \left[ \begin{array}{ccc|c} 
\cdot & \cdot & \cdot & \cdot  \\
\cdot & \cdot & -1 & \cdot \\
\cdot & -1 &   \cdot& \cdot \\ \hline
\cdot &\cdot &\cdot &\cdot 
\end{array}\right]\,,\quad
\widecheck{\beta}^i \!=\!&\, \left[ \begin{array}{ccc|c} 
\cdot & \cdot & \ell_i & \cdot\\
\cdot & \cdot & \cdot &  -e_i \\
\ell_i & \cdot &\cdot & \cdot\\ \hline
\cdot &  f_i &\cdot &\cdot 
\end{array}\right]\,.
\end{align}
For this twisted system the appropriate parity matrix is
\begin{align}
%\label{eq:ConjConjDef}
\label{eq:EtaPrimeDef}
\eta':= &\, -(1\!+\!2(\beta_5)^2)\eta\,,
\end{align} 
with modified conjugate wave-function $\Phibarbar:= \Phi^\dagger\eta'$, 
enabling current bilinears to be constructed in the usual way\footnote{For example, $\Phibarbar \widecheck{\beta}^\mu \Phi$ is a Lorentz four-vector\,.}\,.  Just as in the case of Dirac spinors, it is possible to work with projected wave-functions. For example
$\Phi^{\wedge}:=(-\beta_5^2)\Phi$ is the restriction to the upper
(antisymmetric tensor) components, from which we can further project out self-dual and anti self-dual components as $\pm \ri$ eigenvectors of the chirality matrix $\beta_5$\,
in correspondence with the constraint ${}^*Z_{\mu\nu} = \pm \ri Z_{\mu\nu}$ on the Hodge dual $^*Z_{\mu\nu}:= \textstyle{\frac 12}\varepsilon_{\mu\nu\rho\sigma}
Z^{\rho \sigma}$ of the antisymmetric tensor.
%Comparing (\ref{eq:Beta10Defs}) and (\ref{eq:NewBetaList})\,, it is evident that while for the standard four-vector we have
%\smash{$\eta ( {\beta}{}^\mu ){}^\top \eta=  {\beta}{}^\mu$}\, as required, the rearrangement of blocks in (\ref{eq:NewBetaList}) produces instead the opposite sign,
%\smash{$\eta  (\widecheck{\beta}{}^\mu ){}^\top \eta= -\widecheck{\beta}{}^\mu$}\,.
%However, further conjugation by the Lorentz invariant
%$2(\beta_5)^2+1$ corrects this. The modified parity matrix is then (\ref{eq:EtaPrDef}), namely $\eta':= (2(\beta_5)^2\!+\!1)\eta$, 
%and with the modified conjugate wavefunction
%\begin{align}
%\label{eq:ConjConjDef}
%\Phibarbar:= \Phi^{\dagger} \eta'
%\end{align}
%current bilinears can be constructed in the usual way\footnote{For example, $\Phibarbar \widecheck{\beta}^\mu \Phi$ is a Lorentz four-vector\,.}\,.

It remains to present the variants of the extended DKP formalism which recover the known models for antisymmetric tensor fields (the $KR$ two-form gauge field  and the $AC$ antisymmetric tensor field in symmetric coupling, as well as the 
$CP$ antisymmetric coupling case). In table \ref{tab:DKP_2FGASSTlist} we list the candidate actions in terms of the DKP wave-function, and the corresponding action after
elimination of auxiliary fields which recover the known models. Details of the
calculations are given in appendix \ref{subsec:DKP_2FGAST}, following the method of 
appendix \ref{subsec:DKP_Proca}

\begin{table}[tbp]
%\label{tab:DKP_2FGASSTlist}
\begin{tabular}{|r|c|c|l|}
\hline
&&&\\
& $\mathcal{L}$ (DKP form) &$\Phi$ & $\mathcal{L}$ ($\overline{G}^\mu G_\mu $ form)\\
&&&\\
\hline
&&&\\
$\mathcal{L}^{KR}$ & $\Phibarbar \ri \widecheck{\beta}{}^\mu \partial_\mu\Phi +m \Phibarbar(1+\beta_5^2)\Phi$ & r.(or c.)  & 
$G_\mu=%\textstyle{\frac 12}
\varepsilon_{\mu\nu\rho\sigma}\partial^\nu Z^{\rho\sigma}$\\
&&&\\
\hline
&&&\\
$\mathcal{L}^{AC}$ &$\Phibar\widecheck{\beta}{}^\mu \textstyle{\frac 12}\overleftrightarrow{\partial_\mu} \Phi  + m 
\Phibar\Phi$ & c.s.d. & $G_\mu=\partial^\rho Z_{\rho \mu}$
 \\
&&&\\
\hline
&&&\\
$\mathcal{L}^{CP}$& 
$\textstyle{\frac 12}\big(\Phi^c{}^{\texttt{a}}\beta^\mu \partial_\mu \Phi^{\texttt{b}}\!+\! \ri m \Phibar^\ta \Phi^\tb 
\big)\kappa_{\ta \tb   }\big) \!+\!
\mathrm{h.c.}$&c.s.d & 
$G^\ta_\mu :=\partial^\rho Z^\ta_{\rho\mu}$
\\
&&&\\
\hline
\end{tabular} 
\caption{Extended DKP models (see appendix \ref{subsec:DKP_2FGAST} for derivations).
In each case the DKP action is given along with the type of 10 component wave-function (complex, real, self dual, i.e. $\beta_5\Phi^\wedge = \pm \ri \Phi^\wedge $\, ). The reduced $\overline{G}^\mu G_\mu $ forms after elimination of variables for the two-form gauge, antisymmetric tensor and antisymmetric tensor in antisymmetric coupling are explicitly compared with the standard tensor expressions in the concluding discussion section \ref{sec:Conclusions}.}
\label{tab:DKP_2FGASSTlist}
\end{table}

\section{Lorentz invariant tensor polynomials: potential energy}
\label{sec:Lorentz_potentials}
Given the possibility of antisymmetric tensor fields providing ingredients (together with scalar fields) of an extended symmetry breaking sector, we supplement our count and identification of admissible kinetic terms (for free fields), as in sections \ref{sec:Lorentz_kinetic} and \ref{sec:DKP}\,, with a systematic count of invariant potential terms (both self-coupling interaction and mass terms, where applicable). Invariants are again identified by group theoretical techniques where fields are represented by 
appropriate group characters, and local monomials in fields are classified by character manipulations (with powers in fields counted as symmetrized products).

%In view of the models identified in sections \ref{sec:Lorentz_kinetic} and \ref{sec:DKP}\,, 
Here we work at the $SO(3,1)$ level with complex self-dual antisymmetric tensors
, which are allowed to carry distinct representations of an internal symmetry, taken to be
$SU(n)\times U(1) \cong U(n)$\,, in the fundamental representation, which includes  the standard model for $n=2$, and the Abelian case for $n=1$ .
We also include a complex scalar as proxy for the physical Higgs field.
%Using Young diagram notation for the internal symmetry and the previously introduced $(j_1,j_2)$ notation for 
%
The complex 
self-dual antisymmetric tensor field is now represented as 
\[
%H = \yng{~} + \overline{\yng{~}}\,, \qquad 
(1,0)\yng{~} + (0,1)\overline{\yng{~} }\,,
\]
where the Young diagram stands for a representation of $SU(n)$ with accompanying $U(1)$ eigenvalue the  
diagram weight  $\pm 1$ for the fundamental and contragradient, $\yng{~}$ and $\overline{\yng{~}}$\,, respectively. To include the scalar content we append
\[
(0,0)\yng{~} + (0,0)\overline{\yng{~} }\,,
%Z \cong (1,0)\yng{~} + (0,1)\overline{\yng{~} }\,,\
\]
and seek Lorentz invariants amongst monomials in the complex scalar and tensor at degrees $p,q$ up to quadratic and quartic contributions
(where cross terms with odd powers are disallowed, as the $U(1)$ charge, viz. $(\pm 3)+ (\pm1)$ is non-vanishing).
%\begin{align}
%\label{eq:PowersHT}
%(H+T)^{\uot(2)}=&\, H^{\uot(2)} +  H \cdot T + T^{\uot(2)} \,;\\
%(H+T)^{\uot(4)}=&\, H^{\uot(4)} +  H^{\uot(2)} \cdot T^{\uot(2)} +  T^{\uot(4)} \,.
%\end{align}
%Here the entries on each end implicitly carry $\cdots^{\uot(0)} \equiv 1$ of the omitted character,
%while the cross term in the quartic expansion is now the product of symmetrized second powers
%(see appendix \ref{sec:} for details)\,.
The resulting  count (table \ref{tab:PotentialCount}) lists admissible mass terms for the complex scalar and antisymmetric tensor,
as well as quartic interaction potentials in each, and a possible quartic term with mixed powers
(a bilinear scalar-tensor mixing being excluded). See appendix \ref{subsec:potentials} for details.
\begin{table}[tbp]
\begin{tabular}{|r|cccccc|}
\hline
&&&&&&\\
$p,q$ & 2,0 & 1,1 & 0,2& 4,0& 2,2& 0,4\\
&&&&&&\\
\hline
%\mbox{monomial} & H^{\uot(2)}  & H \!\cdot \!T& T^{\uot(2)}  & H^{\uot(4)} & 
%H^{\uot(2)}\cdot T^{\uot(2)} & T^{\uot(4)} \\
&&&&&&\\
\mbox{count} & \texttt{1} & \texttt{0} & \texttt{0}& \texttt{1} & \texttt{2} & \texttt{1 }\\
&&&&&&\\
\hline
\end{tabular}
\caption{Number of Lorentz invariants for contributions to potential, from monomials 
of the form $(\mathrm{scalar}){}^p\!\cdot\!(\mathrm{antisymmetric\,tensor}){}^q$\,.}
\label{tab:PotentialCount}
\end{table}
The count includes, as expected a single complex scalar mass term and its square, the quartic scalar potential, while forbidding a mass term for the complex self-dual antisymmetric tensor. The quartic pure tensor, and quartic mixed quadratic scalar-tensor, invariants
arise from combinations of the symmetrized contributions of one component of the tensor, say $(1,0)\cdot \yng{~}$\,, with the corresponding opposite weight counterpart(s), in the expansions of either the tensor, or the scalar, respectively. For further details see appendix \ref{subsec:potentials}\,.

\noindent
\section{Conclusions}
\label{sec:Conclusions}
In this work we have provided a robust count of all possible invariant local Lagrangian densities for antisymmetric tensor fields, including both kinetic and potential energy terms, and these have been correlated with different physical models.

While it is straightforward (section \ref{sec:Lagrangian_section}\,) to construct invariants using standard tensor notation for such relativistic fields, we have been at pains
in section \ref{sec:Lorentz_kinetic} to provide a
realization-free identification of the number of admissible linearly independent contributions and their correspondence with known models, which we have also reproduced
in section \ref{sec:DKP} via first order DKP formulations involving an extension to the Kemmer algebra which we believe to be new.
The universality of our ``existence'' count guarantees that, in any alternative presentations such as generalized Bargmann-Wigner, Schwinger,
multispinor or other methods (c.f. \cite{weinberg1964feynman,weinberg1964feynmanII})\,, only and precisely the same
local Lagrangian densities must enter. At the same time, the first order DKP formulation, in which the Kemmer matrix $\beta_5$ plays a central role, analogous to that of $\gamma_5$ in the Dirac equation and spin-$\textstyle{\frac 12}$ context, deepens the analogy between ``chiral Fermions'' and ``chiral Bosons''.

One of the most interesting results of the present paper concerns the identification of the conformally invariant self-dual tensor Lagrangian $\LL^{CP}$\,. This antisymmetric tensor model has been exploited in our recent work \cite{Thierry_Mieg_Jarvis_2021}, in which such ``chiral Bosons'' (c.f. \cite{Chizhov2011}) appear as part of the $SU(2/1)$ superalgebraic extension of the electroweak gauge sector\footnote{In 1979, Ne'eman \cite{ne1979irreducible} and Fairlie \cite{fairlie1979higgs} independently proposed embedding the electroweak Lie algebra of the standard model, $SU(2) \times U(1)$ \,, into the simple Lie-Kac superalgebra $SU(2/1)$ (of which it is the even Lie subalgebra).  
Indeed, the leptons $((\nu_L,e_L)/e_R)$\,,
but also the coloured quark triplets $(u_R /(u_L,d_L)/d_R)$ \cite{dondi1979supersymmetric,neeman1980geometrical}\,, graded by their left/right chirality,
match the lowest irreducible representations of $SU(2/1)\times SU(3)$\,.
Also, we recently presented general results on superalgebra
representations of real and indecomposable type \cite{Thierry_Mieg_Jarvis2023real},
\cite{Jarvis_2022,thierrymieg2023construction}\,, consolidating the original observation \cite{coquereaux1991elementary} that in the case of $SU(2/1)$, the observed three generations of Fermions, associated to the electron, the muon and the tau families, can be naturally accommodated.}\,.
In the work \cite{Thierry_Mieg_Jarvis_2021} we noticed that if such an antisymmetric tensor is coupled to Fermions using the odd matrix generators of $SU(2/1)$ in the appropriate representations, then the Fermion loop counterterms of the quantum field theory induce a new type of tensor propagator, corresponding precisely to $\LL^{CP}$\,.
Moreover, this Lagrangian, being pseudoscalar and antisymmetric in the internal charge space, naturally implements the Landau $CP$ symmetry of the weak interactions.
From the enumeration of Lorentz invariants done in this study, this model indeed occupies the niche allowed by the additional pseudoscalar term, as distinct from the
known scalars (tables \ref{tab:SO31O31Count} and \ref{tab:candidates} above)\,.
This surprising discovery was part of the motivation for the present work.

\vfill
\noindent
\textbf{Acknowledgments:}\\
This research was supported in part by the Intramural Research Program of the National Library of Medicine, National Institute of Health. The authors acknowledge correspondence with Ronald King and Sergei Kuzenko on aspects of this work.
\newpage
%%\nocite{*}
%\bibliographystyle{unsrt}
%%\bibliography{DKP,DKPLit}
%\bibliography{DKP}

\begin{thebibliography}{10}

\bibitem{Dirac1936relativistic}
P.~A.~M. Dirac.
\newblock Relativistic wave equations.
\newblock {\em Proc. Roy. Soc. Lond.}, A155(886):447--459, 1936.

\bibitem{proca1936theorie}
A.~L. Proca.
\newblock Sur la th{{\'e}}orie ondulatoire des {{\'e}}lectrons positifs et
  n{{\'e}}gatifs.
\newblock {\em Journal de Physique et le Radium}, 7(8):347--353, 1936.

\bibitem{fierz1939relativistic}
Markus Fierz and Wolfgang~Ernst Pauli.
\newblock On relativistic wave equations for particles of arbitrary spin in an
  electromagnetic field.
\newblock {\em Proceedings of the Royal Society of London. Series A.
  Mathematical and Physical Sciences}, 173(953):211--232, 1939.

\bibitem{ogievetsky1967notoph}
V.~I. Ogievetsky and I.~V. Polubarinov.
\newblock The notoph and its possible interactions.
\newblock {\em Yad. Fiz.}, 4:216--223, 1967.

\bibitem{KalbRamond1974}
Michael Kalb and P.~Ramond.
\newblock Classical direct interstring action.
\newblock {\em Phys. Rev. D}, 9:2273--2284, Apr 1974.

\bibitem{blau1991topological}
Matthias Blau and George Thompson.
\newblock Topological gauge theories of antisymmetric tensor fields.
\newblock {\em Annals of Physics}, 205(1):130--172, 1991.

\bibitem{Hata1981skew}
T.~Hata, H.~Kugo and N.~Ohta.
\newblock Skew symmetric tensor gauge field theory dynamically realized in
  {QCD} ${U}(1)$ channel.
\newblock {\em Nuclear Physics. B}, 178(3):527, 1981.

\bibitem{Choi_2015}
Kang-Sin Choi and Jeong-Hyuck Park.
\newblock Standard model as a double field theory.
\newblock {\em Physical Review Letters}, 115(17), oct 2015.

\bibitem{AvdeevChizhov1994}
L.~V. Avdeev and M.~V. Chizhov.
\newblock Antisymmetric tensor matter fields. {A}n abelian model.
\newblock {\em Phys. Lett. B}, 321:212--18, 1994.

\bibitem{Chizhov2011}
M.~V. Chizhov.
\newblock Theory and phenomenology of spin-1 chiral particles.
\newblock {\em Phys. Part. Nuc.}, 42:93--183, 2011.

\bibitem{LemesRenanSorella1995II}
V.~Lemes, R.~Renan, and S.~P. Sorella.
\newblock $\varphi_{4}^{4}$-theory for antisymmetric tensor matter fields in
  {M}inkowski space-time.
\newblock {\em Phys. Lett. B}, 352:37--42, 1995.

\bibitem{wetterich2008quantization}
C.~Wetterich.
\newblock Quantization of chiral antisymmetric tensor fields.
\newblock {\em International Journal of Modern Physics A}, 23(10):1545--1579,
  2008.

\bibitem{Thierry_Mieg_Jarvis_2021}
Jean Thierry-Mieg and Peter~D. Jarvis.
\newblock {SU}(2/1) superchiral self-duality: a new quantum, algebraic and
  geometric paradigm to describe the electroweak interactions.
\newblock {\em Journal of High Energy Physics}, 2021(4), apr 2021.

\bibitem{duffin1938characteristic}
R.~J. Duffin.
\newblock On the characteristic matrices of covariant systems.
\newblock {\em Physical Review}, 54(12):1114, 1938.

\bibitem{Kemmer1939}
N.~Kemmer.
\newblock The particle aspect of meson theory.
\newblock {\em Proc. Roy. Soc. A}, 173:91--116, 1939.

\bibitem{petiau1938university}
G.~Petiau.
\newblock University of {P}aris thesis (1936). {P}ublished in: {A}cad. {R}oy.
  de {B}elg., {C}lasse sci., {M}em in 8o 16, no. 2 (1936).

\bibitem{Thierry_Mieg_2020}
Jean Thierry-Mieg.
\newblock Scalar anomaly cancellation reveals the hidden superalgebraic
  structure of the quantum chiral {SU}(2/1) model of leptons and quarks.
\newblock {\em Journal of High Energy Physics}, 2020(10), Oct 2020.

\bibitem{Thierry_Mieg_2021}
Jean Thierry-Mieg.
\newblock Chirality, a new key for the definition of the connection and
  curvature of a {L}ie-{K}ac superalgebra.
\newblock {\em Journal of High Energy Physics}, 2021(1), Jan 2021.

\bibitem{bateman1910transformation}
Harry Bateman.
\newblock The transformation of the electrodynamical equations.
\newblock {\em Proceedings of the London Mathematical Society}, 2(1):223--264,
  1910.

\bibitem{ThierryMiegJarvis2023conformal}
Jean Thierry-Mieg and Peter~D. Jarvis.
\newblock Conformal invariance of antisymmetric tensor field theories in any
  even dimension.
\newblock {\em Preprint arXiv:2311.01701v2 [hep-th]}, 2023.

\bibitem{KrajcikNieto1977historical}
R.~A. Krajcik and Michael~Martin Nieto.
\newblock Historical development of the {B}habha first-order relativistic wave
  equations for arbitrary spin.
\newblock {\em American Journal of Physics}, 45(9):818--822, 1977.

\bibitem{NedjadiBarrett1993}
Y.~Nedjadi and R.~C. Barrett.
\newblock On the properties of the {D}uffin-{K}emmer-{P}etiau equation.
\newblock {\em J. Phys. G: Nucl. Part. Phys.}, 19:87--98, 1993.

\bibitem{NedjadiBarrett1994}
Y.~Nedjadi and R.~C. Barrett.
\newblock {Solution of the central field problem for a Duffin--Kemmer--Petiau
  vector boson}.
\newblock {\em Journal of Mathematical Physics}, 35(9):4517--4533, 09 1994.

\bibitem{Kruglov2010kalb}
S.~I. Kruglov.
\newblock Kalb--{R}amond fields in the {P}etiau-{D}uffin-{K}emmer formalism and
  scale invariance.
\newblock {\em Modern Physics Letters A}, 25(32):2745--2751, October 2010.

\bibitem{kruglov2011field}
S.~I. Kruglov.
\newblock Field theory of massive and massless vector particles in the
  {D}uffin-{K}emmer-{P}etiau formalism.
\newblock {\em International Journal of Modern Physics A}, 26(15):2487--2501,
  June 2011.

\bibitem{BeltranPimentelSoto2020}
J.~Beltran, B.~M. Pimentel, and D.~E. Soto.
\newblock The causal approach proof for the equivalence of {SDKP}$_{4}$ and
  {SQED}$_{4}$ at tree{-}level.
\newblock {\em Mod. Phys. A}, 35:2050042, 2020.

\bibitem{weinberg1964feynman}
Steven Weinberg.
\newblock Feynman rules for any spin.
\newblock {\em Physical Review}, 133(5B):B1318, 1964.

\bibitem{weinberg1964feynmanII}
Steven Weinberg.
\newblock Feynman rules for any spin. {II}. massless particles.
\newblock {\em Physical Review}, 134(4B):B882, 1964.

\bibitem{ne1979irreducible}
Yuval Ne'eman.
\newblock Irreducible gauge theory of a consolidated {S}alam-{W}einberg model.
\newblock {\em Physics Letters B}, 81(2):190--194, 1979.

\bibitem{fairlie1979higgs}
D.~B. Fairlie.
\newblock Higgs fields and the determination of the {W}einberg angle.
\newblock {\em Physics Letters B}, 82(1):97--100, 1979.

\bibitem{dondi1979supersymmetric}
P.~H. Dondi and Peter~D. Jarvis.
\newblock A supersymmetric {W}einberg-{S}alam model.
\newblock {\em Physics Letters B}, 84(1):75--78, 1979.

\bibitem{neeman1980geometrical}
Yuval Ne'eman and Jean Thierry-Mieg.
\newblock Geometrical gauge theory of ghost and {G}oldstone fields and of ghost
  symmetries.
\newblock {\em Proceedings of the National Academy of Sciences},
  77(2):720--723, 1980.

\bibitem{Thierry_Mieg_Jarvis2023real}
Jean Thierry-Mieg and Peter~D. Jarvis.
\newblock New {H}ermitian conjugations and real forms in $sl(m/n)$ and
  $osp(2/2n)$ superalgebras modelling quark and lepton families.
\newblock {\em {P}reprint}, 2023.

\bibitem{Jarvis_2022}
Peter~D. Jarvis and Jean Thierry-Mieg.
\newblock Indecomposable doubling for representations of the type {I} {L}ie
  superalgebras $sl(m/n)$ and $osp(2/2n)$.
\newblock {\em Journal of Physics A: Mathematical and Theoretical},
  55(47):475206, {N}ov 2022.

\bibitem{thierrymieg2023construction}
Jean Thierry-Mieg, Peter~D. Jarvis, and Jerome~Germoni with an appendix~by
  Maria~Gorelik.
\newblock Construction of matryoshka nested indecomposable {$N$}-replications
  of {K}ac-modules of quasi-reductive {L}ie superalgebras, including the
  $sl(m/n)$ and $osp(2/2n)$ series.
\newblock {\em Preprint arXiv:2207.06538v4 [math.RT]}, 2023.

\bibitem{coquereaux1991elementary}
Robert Coquereaux.
\newblock Elementary fermions and $su(2|1)$ representations.
\newblock {\em Physics Letters B}, 261(4):449--458, 1991.

\bibitem{fulling1992normal}
S.~A. Fulling, Ronald~C. King, B.~G. Wybourne, and C.~J. Cummins.
\newblock Normal forms for tensor polynomials. {I}. the {R}iemann tensor.
\newblock {\em Classical and Quantum Gravity}, 9(5):1151, 1992.

\bibitem{king1982some}
R.~C. King and B.~G. Wybourne.
\newblock Some noteworthy spin plethysms.
\newblock {\em Journal of Physics A: Mathematical and General}, 15(4):1137,
  1982.

\bibitem{BlackKingWybourne83}
G.~R.~E. Black, R.~C. King, and B.~G. Wybourne.
\newblock Kronecker products for compact semisimple {L}ie groups.
\newblock {\em Journal of Physics A: Mathematical and General}, 16(8):1555,
  1983.

\bibitem{LemesRenanSorella1995I}
V.~Lemes, R.~Renan, and S.~P. Sorella.
\newblock New algebraic renormalization of antisymmetric tensor matter fields.
\newblock {\em Phys. Lett. B}, 344:158--63, 1995.

\bibitem{Jackiw_2011}
R.~Jackiw and S-Y. Pi.
\newblock Tutorial on scale and conformal symmetries in diverse dimensions.
\newblock {\em Journal of Physics A: Mathematical and Theoretical},
  44(22):223001, {M}ay 2011.

\end{thebibliography}

\vfill
%\pagebreak
\begin{appendix}
\section{Lorentz invariant tensor polynomials}
\label{sec:TensorPoly}
\subsection{Kinetic energy}
\label{subsec:kinetic}
\mbox{}\\

With the notation $[ \cdots ]^{\uot(2)}, [ \cdots ]^{\uot(1^2)} $ 
to represent the symmetric or antisymmetric symmetrized quadratic tensor power,
our task is to count $SO(3,1)$ invariants in the expansion of equation (\ref{eq:KineticCharacter}), namely
\begin{equation}
\big( (0,0)\!+\!(1,1)\big)\!\cdot\!{[}(1,0)\!+\!(0,1){]}^{\uot(2)}\,+
\big( (0,0)\!+\!(1,1)\big)\!\cdot\! {[}(1,0)\!+\!(0,1){]}^{\uot(1^2)}\,.
\end{equation}
%respectively:
%\begin{align}
%\label{eq:SO31tasksA}
%SO(3,1)_{\texttt{S}} :\qquad&\, \big( (0,0)+(1,1)\big)\cdot {[}(1,0)+(0,1){]}^{\uot(2)}\,;\\
%SO(3,1)_{\texttt{A}}:\qquad&\, \big( (0,0)+(1,1)\big)\cdot {[}(1,0)+(0,1){]}^{\uot(1^2)}\,.
%\nonumber
%\end{align}
Using the standard rules for symmetrization of rotation group representations, viz. $(1)^{\uot(2)} =(0)+(2) $\,,
$(1)^{\uot(1^2)} =(1) $\,, and the fact that  the symmetrization or antisymmetrization
of a sum of two parts contains a single copy of the cross term, we require
$SO(3,1)$ invariants in 
\[
\big( (0,0)\!+\!(1,1)\big)\!\cdot\!{[}(0,0) \!+\!(2,0)+(0,0)+(0,2)+(1,1){]}\, \quad
\mbox{and} \quad
\big( (0,0)\!+\!(1,1)\big)\!\cdot\! {[}(1,0)\!+\!(0,1)\!+\!(1,1){]}\,,
\]
%
%\begin{align}
%\label{eq:SO31tasksB}
%SO(3,1)_{\texttt{S}} :\qquad&\, \big( (0,0)+(1,1)\big)\cdot 
%{[}(0,0) +(2,0)+(0,0)+(0,2)+(1,1){]}\,;\\
%SO(3,1)_{\texttt{A}}:\qquad&\, \big( (0,0)+(1,1)\big)\cdot {[}(1,0)+(0,1)+(1,1){]}\,.\nonumber
%\end{align}
and by matching we find three invariants in symmetric, and one in antisymmetric
coupling, respectively:
%\begin{align}
%\label{eq:SO31tasksC}
%SO(3,1)_{\texttt{S}} :\qquad&\, \Rightarrow \quad 3(0,0) \,\quad  \mbox{ (3 invariants)}\,;\\
%SO(3,1)_{\texttt{A}}:\qquad&\, \Rightarrow\quad  1(0,0) \, \quad \mbox{(1 invariant)}\,.\nonumber
%\end{align}
\begin{align}
\label{eq:SO31Count}
\begin{array}{c |c  c}
\mbox{$SO(3,1)$} &
\underbrace{\big( (0,0)\!+\!(1,1)\big)\!\cdot\!{[}(1,0)\!+\!(0,1){]}^{\uot(2)}} & 
\underbrace{\big( (0,0)\!+\!(1,1)\big)\!\cdot\! {[}(1,0)\!+\!(0,1){]}^{\uot(1^2)}}
\\
\mbox{count} &  3 & 1
\end{array}
\end{align}
A short cut to the needed classification of terms under the full Lorentz group is provided by working initially with tensors under the larger structure group $GL(4)$\,, and restricting to $O(3,1)$ in the final step. In this case the kinetic term 
is now an irreducible symmetric second rank tensor, and the antisymmetric tensor 
field simply an irreducible antisymmetric second rank tensor, represented as $\yng{~~}$ and
$\yng{~,~}$\,, respectively\footnote{The computation is aided visually 
by use of Young diagram notation for tensor manipulation (see for example \cite{fulling1992normal})\,.}. The counting needed is thus to identify $O(3,1)$ invariants in the reduction
of the $GL(4)$ characters
\[
%\label{eq:GL4tasks}
\yng{~~}\!\cdot \!\yng{~,~}^{\uot(2)}\,\quad \mbox{and} \quad
\yng{~~}\!\cdot \! \yng{~,~}^{\uot(1^2)}\,, 
\]
respectively.
%\begin{align}
%\label{eq:O31tasks}
%O(3,1)_{\texttt{S}}:\qquad &\, \Rightarrow \quad 2(0,0)+(0,0)^* \,\quad  \mbox{ (2 scalar, 1 pseudoscalar)}\,;\\
%O(3,1)_{\texttt{A}}:\qquad&\, \Rightarrow \quad (0,0)^* \, \quad \mbox{(1 pseudoscalar)}\,.
%\end{align}
%\begin{align}
%\label{eq:GL4O31Count}
%\begin{array}{r |c  c}
%GL(4) &
% {\yng{~~}\!\cdot \!\yng{~~}^{\uot(2)}} & 
%{\yng{~~}\!\cdot \! \yng{~,~}^{\uot(1^2)}}
%\\[.2cm]
%\mbox{$O(3,1$ scalar} &  2 & 0 \\
%\mbox{$O(3,1)$ pseudoscalar} &  1 & 1 \\
%\end{array}
%\end{align}
%
%The count of candidate kinetic terms for antisymmetic tensor fields forming one dimensional representations under the full Lorentz group proceeds at the $GL(4)$ level by reducing the symmetric and antisymmetric products (section \ref{sec:Lorentz_kinetic})
%\[
%%\label{eq:GL4tasks}
%\yng{~~}\!\cdot \!\yng{~,~}^{\uot(2)}\,\quad \mbox{and} \quad
%\yng{~~}\!\cdot \! \yng{~,~}^{\uot(1^2)}\,. 
%\]
Given that
\[
\yng{~,~}^{\uot(2)} =\yng{~,~,~,~} + \yng{~~,~~} \,, \quad \mbox{and}\quad  
\yng{~,~}^{\uot(1^2)} = \yng{~~,~,~}  \,,
\]
we look for orthogonal group\footnote{We use combinatorial character manipulation methods with $O(4)$ diagrams
being a proxy for (finite dimensional) tensor reps. We have 
$\yng{~} \cong \underline{4}$\,, $\yng{~,~} \cong \underline{6}$. Under 
$O(3,1)\downarrow SO(3,1)$\,, two-rowed diagrams reduce to 
inequivalent irreducible parts, viz. 
$\yng{~,~} \cong \yng{~,~}_+ + \yng{~,~}_-$ (self-dual and anti self-dual tensors, respectively).
For $\yng{~~~,~~}\cong \yng{~~~,~~}_+ + \yng{~~~,~~}_-$ etc.\,, with row lengths ${[}\ell_1,\ell_2{]}$\,, the rep is irreducible if $\ell_2=0$ and reduces to $SO(3,1)$
$(\textstyle{\frac 12}\ell_1, \textstyle{\frac 12}\ell_1)$, and for $\ell_1 \ge \ell_2 > 0$ 
the decomposition is $(\textstyle{\frac 12}(\ell_1 +\ell_2), \textstyle{\frac 12}(\ell_1 -\ell_2)) + (\textstyle{\frac 12}(\ell_1 -\ell_2), \textstyle{\frac 12}(\ell_1 +\ell_2))$\,.} invariants in the general linear group products
\footnote{Diagrams with five or more rows have been removed because the corresponding $GL(4)$ characters are zero.} 
\begin{align*}
\yng{~~}.\yng{~~,~~} + \yng{~~}.\yng{~,~,~,~} = &\,
  \yng{~~~~,~~}+\yng{~~~,~~,~}+\yng{~~,~~,~~}+ \yng{~~~,~,~,~}\\ 
%+ \cancelto{ }{\yng{~~,~,~,~,~}}\,;\\
\mbox{and}\quad
\yng{~~}.\yng{~~,~,~}= &\,\yng{~~~~,~,~} +\yng{~~~,~~,~}+\yng{~~~,~,~,~}+\yng{~~,~~,~,~}\,.
%\yng{~,~}.\yng{~~,~~} + \yng{~,~}.\yng{~,~,~,~} =&\,
%\yng{~~~,~~~}+ \yng{~~~,~~,~}+\yng{~~,~~,~,~}\,.
%+\cancelto{ }{\yng{~,~,~,~,~,~}}+ %\cancelto{ }{\yng{~~,~,~,~,~}}+\yng{~~,~~,~,~}\,.
\end{align*}
Reduction to the orthogonal group requires diagram manipulation equivalent to removing all possible contractions with the metric tensor. In symmetric coupling
we have firstly
\begin{align*}
\yng{~~~~,~~} \Rightarrow  &\, \yng{~~~~,~~}+\yng{~~~~}+\yng{~~~,~}+\yng{~~,~~} +2.\yng{~~} +\scalebox{.6}{$\bullet$} \, ,\\
(126 = &\, 42+25+30+10+18+1)\,;\\
\yng{~~,~~,~~} \Rightarrow  &\, \yng{~~,~~,~~} + \yng{~~,~~}+ \yng{~~}+  \scalebox{.6}{$\bullet$}\Rightarrow - \yng{~~,~~} + \yng{~~,~~} +\yng{~~}+   \scalebox{.6}{$\bullet$}\,,\\
(10= &\, 9+1) \,,
\end{align*}
where ``\scalebox{.6}{$\bullet$}'' stands for the scalar (one dimensional) representation, and dimension checks have been included. Note that the second expansion contains a non-standard orthogonal group diagram (with more than two rows) which modifies to the negative of an irreducible character\footnote{If there are $2+h$ rows the modification is by removal of  boundary strip of length $2h$, starting with the offending rows, with accompanying sign reversal, or change  to an associated representation 
(``\,${}^*$\,''), or removal if an improper diagram results. For notation and methods see \cite{king1982some},\cite{BlackKingWybourne83}\,.}\,.

From the above diagrams it is of note that one $GL(4)$ term occurs in  both 
symmetric and antisymmetric coupling, corresponding to two distinct 
(linearly independent) couplings. The orthogonal reduction follows
\begin{align*}
\yng{~~~,~,~,~} \Rightarrow  &\, \yng{~~~,~,~,~}+ \yng{~,~,~,~} + \yng{~~,~,~} 
\,\Rightarrow  \quad \scalebox{.6}{$\bullet$}^* + \yng{~~}^*\\
(10= &\, 1+9) \,,
\end{align*}
where the modification rules have been applied (with removal of a vanishing character).
This term therefore supplies the pseudo-scalar candidate 
$\scalebox{.6}{$\bullet$}^*$ in both coupling symmetries.

Verifying that the remaining $GL(4)$ terms do not contain further scalars or pseudoscalars completes the accounting of kinetic terms admissible 
under the full Lorentz group, with the results (c.f. equation (\ref{eq:SO31Count}) and table \ref{tab:SO31O31Count}):
\begin{align}
\label{eq:GL4O31Count}
\begin{array}{r |c  c}
GL(4) &
 {\yng{~~}\!\cdot \!\yng{~~}^{\uot(2)}} & 
{\yng{~~}\!\cdot \! \yng{~,~}^{\uot(1^2)}}
\\[.2cm]
\mbox{$O(3,1)$ scalar} &  2 & 0 \\
\mbox{$O(3,1)$ pseudo-scalar} &  1 & 1 \\
\end{array}
\end{align}
% (equation (\ref{eq:GL4O31Count}) above; c.f. 
%equation (\ref{eq:SO31Count})\,.) 

\subsection{Potential energy}
\label{subsec:potentials}
\mbox{}\\
The count of admissible local invariant interaction terms for the complex scalar and self-dual antisymmetric tensor fields (section \ref{sec:Lorentz_potentials}) by group methods requires working with
expansions of symmetrized powers of the characters representing them.  
Here we provide more details for the quartic couplings (the scalar quartic is the square of the quadratic mass term).
As already noted, the the requirement to have only 
monomials with net vanishing $U(1)$ charge
considerably restricts the count. For example 
at quartic degree\,, the only possible contributions from the antisymmetric tensor field come from the nine terms in
\begin{align*}
%\label{eq:T4admissible}
[(1,0)\!\cdot\!\yng{~}]^{\uot{(2)}} \!\cdot\! [(1,0)\cdot \overline{\yng{~}}]^{\uot{(2)}} =&\,
[(0+2,0)\!\cdot\!\yng{~~} + (1,0)\!\cdot\!\yng{~,~}]\!\cdot\!
[(0,0\!+\!2)\!\cdot\!\overline{\yng{~~}} +  (0,1)\!\cdot\!\overline{\yng{~,~}}]\\
=&\, (0\!+\!2,0\!+\!2)\!\cdot\! \yng{~~}\!\cdot\!\overline{\yng{~~}} + 
(0\!+\!2,1)\!\cdot\!\yng{~~}\!\cdot\! \overline{\yng{~,~}} + 
(1,0\!+\!2)\!\cdot\!\overline{\yng{~~}}\!\cdot\! {\yng{~,~}} 
+ (1,1)\yng{~,~}\!\cdot\! \overline{\yng{~,~}}\,.
\end{align*}
wherein the $(0,0)\!\cdot\!\yng{~~}\!\cdot\! \overline{\yng{~~}}$ term contains a unique
one dimensional representation\footnote{Compound diagrams provide a succinct way to handle contragradient representations \cite{BlackKingWybourne83}\,.}. On the other hand for the mixed quadratic scalar-tensor term 
%$H^{\uot(2)}\!\cdot \!T^{\uot(2)}$\,, 
we have
\begin{align*}
[\yng{~~}+\yng{~}\!\cdot\!\overline{\yng{~}} + \overline{\yng{~~}}]\!\cdot\!
[(0+2,0)\!\cdot\!\yng{~~} + (1,1)\!\cdot\!\yng{~}\!\cdot\!\overline{\yng{~}}
+ (0,0+2)\!\overline{\yng{~~}}]
\end{align*}
wherein two invariants will come from the expansions of
\[
\yng{~~}\cdot\!(0,0)\!\cdot\!\overline{\yng{~~}}\,
\quad \mbox{and}\quad
\overline{\yng{~~}}\!\cdot\!(0,0)\!\cdot\!\yng{~~}\,.
\]
Note that the present count
(table 
\ref{tab:PotentialCount}) holds for internal symmetry $U(n) \cong SU(n)\times U(1)$
which includes the standard model $SU(2)\times U(1)$ and also the Abelian case of complex fields.

\noindent
\section{The DKP algebra} 
\label{sec:KemmerAlgebra}
\mbox{}\\
For completeness, we give explicit representations of the Kemmer $\beta$-matrices for (in $1\!+\!3\!+\!1$ block form) the $5\times 5$\,: 
\begin{align*}
\beta^0=&\,\left[ \begin{array}{c|c|c} 
\cdot & \cdot & 1\\ \hline
\cdot & \cdot & \cdot \\ \hline
1 & \cdot & \cdot \\
\end{array}\right]\,, \quad
\beta^i=\left[ \begin{array}{c|c|c} 
\cdot & \cdot & \cdot \\ \hline
\cdot & \cdot & e_i  \\ \hline
\cdot & -f_i & \cdot \\
\end{array}\right]\,;
\end{align*}
and (in $3\!+\!3\!+\!3\!+\!1$ block form) $10\times 10$\,: 
\begin{align*}
\beta^0 \!=\! \left[ \begin{array}{ccc|c} 
\cdot & \cdot & 1 & \cdot  \\
\cdot & \cdot & \cdot & \cdot \\
1 & \cdot &   \cdot& \cdot \\ \hline
\cdot &\cdot &\cdot &\cdot 
\end{array}\right]\,,\quad
\beta^i \!=\!&\, \left[ \begin{array}{ccc|c} 
\cdot & \cdot & \cdot & \texttt{a} e_i\\
\cdot & \cdot & \texttt{b}\ell_i & \cdot \\
\cdot & \texttt{b}'\ell_i &\cdot & \cdot\\ \hline
\texttt{a}' f_i &\cdot &\cdot &\cdot 
\end{array}\right]\,;\quad
\end{align*}
representations\,. Here $e_i$\,, $i=1,2,3$ are standard 3-component unit column vectors, $f_i = \smash{{}^\top e_i}$ the corresponding row vectors, and $\ell_i :=
\textstyle{\frac 12}\varepsilon_{ijk} \ell_{jk}$ elementary $3\times 3$ rotation generators\footnote{Where $\ell_{jk} := e_{jk}-e_{kj}= e_j\otimes f_k - e_k\otimes f_j$\,, with commutation relations ${[}\ell_i, \ell_j{]} = -\varepsilon_{ijk}\ell_k$ and matrix elements $(\ell_i)_{jk} =\varepsilon_{ijk}$\,, and for example $\ell_ie_i =0$\,,
$(\ell_i)^2 = e_{i i}-1$\,.}.
Examination of the form of the $SO(3,1)$ rotation
generators ${[}\beta_i,\beta_j{]}$\,, the boost generators ${[}\beta_0,\beta_i{]}$\,,
as well as algebra identities such as
${\{} \beta^j,(\beta^i)^2{\} } = -\beta^j$\,, ${\{} (\beta^i)^2,(\beta^0){\} } = -\beta^0$
yields $\texttt{a}\texttt{a}'=-1$\,, $\texttt{b}\texttt{b}' = +1$\,. 
Hereafter we choose $\texttt{a}=\texttt{b}=+1$\,.

For the pseudo-scalar $\beta_5 := 
\textstyle{\frac 12}\varepsilon^{\mu \nu\rho\sigma}
\beta_\mu\beta_\nu\beta_\rho\beta_\sigma$ direct evaluation gives after normalization
\begin{align*}
{\beta_5} \!=\! \left[ \begin{array}{ccc|c} 
\cdot & 1 & \cdot & \cdot  \\
-1& \cdot & \cdot & \cdot \\
\cdot & \cdot &   \cdot& \cdot \\ \hline
\cdot &\cdot &\cdot &\cdot 
\end{array}\right]\,,\quad
\end{align*}
from which we derive the 10-component twisted  $\widecheck{\beta}_\mu = {[}\beta_5,\beta_\mu{]}$ matrices\,,
\begin{align*}
\widecheck{\beta}^0 \!=\! \left[ \begin{array}{ccc|c} 
\cdot & \cdot & \cdot & \cdot  \\
\cdot & \cdot & -1 & \cdot \\
\cdot & -1 &   \cdot& \cdot \\ \hline
\cdot &\cdot &\cdot &\cdot 
\end{array}\right]\,;\quad
\widecheck{\beta}^i \!=\!&\, \left[ \begin{array}{ccc|c} 
\cdot & \cdot & \ell_i & \cdot\\
\cdot & \cdot & \cdot &  -e_i \\
\ell_i & \cdot &\cdot & \cdot\\ \hline
\cdot & f_i &\cdot &\cdot 
\end{array}\right]\,.
\end{align*}

The parity operators are defined
\begin{align}
\label{eq:EtaEta'Def}
\eta:= &\,2\beta_0^2-1 = \mathrm{diag}(1,-1,1,-1)\,, \quad \nonumber\\
\eta':= &\,-(1+2\beta_5^2)\eta =\mathrm{diag}(1,-1,-1,1)\,, \\
\mbox{with}\quad
\eta { {\beta}{}^\mu}^\top \eta = &\, +  {\beta}{}^\mu=-\eta' {{\beta}{}^\mu}^\top \eta' \,,\quad
\eta' {\widecheck{\beta}{}^\mu}^\top \eta' =  + \widecheck{\beta}{}^\mu=
-\eta { \widecheck{\beta}{}^\mu}^\top \eta\,, \nonumber
%&\, -  {\beta}{}^\mu\,,
%\qquad 
%\eta {\widecheck{\beta}{}^\mu}^\top \eta =  - \widecheck{\beta}{}^\mu\, \nonumber
\end{align}
from which we define $\Phibar := \Phi^\dagger \eta$ and
$\Phibarbar := \Phi^\dagger \eta'$\,. Sundry properties which we shall use are
\begin{align}
\label{eq:BetaProps}
\{\beta_5^2,\beta^\mu\} =&\,-\beta^\mu\,,
\quad \{\beta_5^2,\widecheck{\beta}{}^\mu\} =-\widecheck{\beta}{}^\mu\,;\nonumber\\
\beta_\mu \beta^\mu =&\, 3+\beta_5^2\,, \quad
\widecheck{\beta}_\mu \beta^\mu = 2\beta_5^2\,, \\
(\beta_5)^3 =&\, -\beta_5\,, \quad {\beta}_\mu \beta^\sigma\beta^\mu = 2\beta^\sigma\,. \nonumber
\end{align}

%
%As explained in the main text, the introduction of the $\beta_5$ twist is responsible for
%switching the standard identification of wavefunction components, from the
%conventional DKP analogue of the Proca (massive vector) system, over to a 
%DKP analogue of the Adveev-Chishov antisymmetric tensor matter field system. 

\section{Extended DKP formalism}
\label{sec:DKPapp}

\subsection{Proca massive vector field}
\label{subsec:DKP_Proca}
As a case study, and to illustrate the reduction of first order DKP actions to standard
forms, we reproduce here the standard treatment for the derivation of the
Proca (complex massive spin-1) system. In an obvious notation we have
\begin{align*}
\Phibar(\ri\beta^\mu \partial_\mu \Phi )-m\Phibar\Phi =&\,
(\uoE, -\uoB,\uoA,-A_0)\left[ \begin{array}{ccc|c} 
-m & \cdot & \partial_0 & \unab  \\
\cdot & -m & \uell\!\cdot\!\unab & \cdot  \\
\partial_0 & \uell\!\cdot\!\unab  & -m & \cdot   \\ \hline
-\unab\!\cdot &\cdot &\cdot & -m
\end{array}\right]\left(\begin{array}{c}
\uE \\ \uB \\\uA \\ A_0\end{array}\right)\,.\\
%=&\,\uoE\cdot(\ri \partial_0 \uA +\ri\unab A_0) +
%(-\uoB)\cdot(  -\ri\unab \times \uA ) +
% \uoA\cdot( \ri \partial_0 \uE -\ri \unab \times \uB  ) 
%+(-\oA_0)(-\ri \unab\cdot \uE )\\
%&\, - M(\uoE\cdot\uE - \uoB\cdot\uB)
%-M(\uoA\cdot\uA - \oA_0A_0)\,.\\
= &\,\uoE\cdot(\ri \partial_0 \uA +\ri\unab A_0) +
\uE\cdot (-\ri \partial_0 \uoA -\ri\unab \oA_0)
+ \uoB\cdot(\ri\unab \times \uoA) + \uB\cdot (-\ri\unab \times \uoA)+ \\
&\, - m(\uoE\cdot\uE - \uoB\cdot\uB)
-m(\uoA\cdot\uA - \oA_0A_0)
\end{align*}
%Reordering terms up to total derivatives and eliminating $\uoE,\uE, \uoB,\uB$ using their (algebraic) equations of motion yields
The expansion in three-vector calculus involving $\uE, \uB$ and $\uA$ (with scalar $A_0$) and complex conjugates follows using
matrix forms such as $(\ul \cdot \unab ) \Rightarrow -\unab \times $\, and 
use of partial integration. After eliminating auxiliary fields $\uE,\uB$ using their equations of motion we find
\begin{align*}
\Phibar(\ri\beta^\mu \partial_\mu \Phi )-m\Phibar\Phi 
%= &\,\uoE\cdot(\ri \partial_0 \uA +\ri\unab A_0) +
%\uE\cdot (-\ri \partial_0 \uoA -\ri\unab \oA_0)
%+ \uoB\cdot(\ri\unab \times \uoA) + \uB\cdot (-\ri\unab \times \uoA)+ \\
%&\, - m(\uoE\cdot\uE - \uoB\cdot\uB)
%-m(\uoA\cdot\uA - \oA_0A_0)\\
\Rightarrow &\,
+\frac{1}{m}( \partial_0 \uoA +\unab \oA_0)\cdot ( \partial_0 \uA +\unab A_0)
-\frac{1}{m}(\unab \times \uoA)\cdot(\unab \times \uA) +m(\oA_0A_0-\uoA\cdot\uA) 
\end{align*}
which after rescaling $A_0 \rightarrow -\sqrt{m}A_0$\,, $\uA \rightarrow \sqrt{m}\uA$
accords with the free action for a complex vector field of mass $m$\,,
\[
\mathscr{L}_{Proca} = -\textstyle{\frac 12} \uF^{\mu\nu}F_{\mu\nu} + m^2\oA^\mu A_\mu\,
\]
with field strength $F_{\mu\nu}= \partial_\mu A_\mu-\partial_\nu A_\mu$ and the
standard identification of electric and magnetic fields
\footnote{
Elimination of auxiliary fields amounts to changing the normalization of  the generating function by a functional Gaussian. 
From the above expansion, it is evident that elimination of $A_0, \uA $ instead of $ \uE,  \uB$ would lead to a `dual' model equivalent to the Proca theory, at least at the non-interacting level.}  $\uE, \uB$\,.

\subsection{Two-form gauge potentials and antisymmetric tensor fields}
\label{subsec:DKP_2FGAST}
In the sequel we adopt the following notation for fields in the 10 component wave-function of the modified DKP systems:
\begin{align}
\label{eq:ASTLabels}
\Phi = {}^\top\big( \uS, \uR,\uG,G_0\big)\,,
\quad \Phibar = (\uSb, -\uRb,  \uGb, -\Gb_0)\,,\,,\quad
\Phibarbar = (\uSb, -\uRb, -\uGb, +\Gb_0)\,,
\end{align}
together with $\Phi^c := {}^\top\Phi \eta$\,, in which we identify the three-vector $\uS=(Z_{01},Z_{02},Z_{03})$\,, and pseudo-vectors $\uR=(Z_{23},Z_{31},Z_{12})$\, as putative ``electric'' and ``magnetic '' parts of an antisymmetric tensor 
$Z_{\mu\nu}$, the Hodge dual of whose three-form curl  (in the twisted case) becomes the axial vector field strength
$G_\mu \cong (G_0,\uG)$\, consisting of a pseudo-vector together with a pseudoscalars.
The allocations in (\ref{eq:ASTLabels}) to the components of $Z_{\mu\nu}$
in order to reassemble the DKP-derived actions as kinetic terms in local relativistic invariants in standard tensor notation, follow after field rescaling.
 
\noindent
\textbf{Case 1: two-form gauge field \cite{ogievetsky1967notoph,KalbRamond1974}:}\mbox{}\\
Following the above discussion we consider 
\begin{align*}
\Phibarbar \ri \widecheck{\beta}{}^\mu \partial_\mu\Phi +m \Phibarbar(1+\beta_5^2)\Phi =&\, \\
= 
\uoG\cdot(\ri \partial_0 \uR + \ri\unab\times\uS)  +&\,\oG_0(-\ri\unab\cdot \uR )
+\uG\cdot(-\ri \partial_0 \uoR -\ri  \unab\times \uoS ) 
+G_0(+\ri\unab\cdot \uoR           ) +m(\oG_0 G_0-\uoG\dot \uG)\\
\Rightarrow &\,
%-(-\ri \partial_0 \uoR -\ri  \unab\times \uoS ) \cdot(\ri \partial_0 \uR + \ri\unab\times\uS)+ (+\ri\unab\cdot \uoR)( -\ri\unab\cdot \uR  )\\
%\equiv &\,
\frac{1}{m}\left(
(\partial_0 \uoR +\unab\times \uoS )\cdot
(\partial_0 \uR +  \unab\times\uS)-(\unab\cdot \uoR)(\unab\cdot \uR)
\right)
\end{align*}
so finally with $\uR, \uS \rightarrow -\sqrt{m}\uR, \sqrt{m}\uS$\,,
\[
\Phibarbar \ri \widecheck{\beta}{}^\mu \partial_\mu\Phi +m \Phibarbar(1+\beta_5^2)\Phi  \Rightarrow
\left((-\unab\cdot \uoR)(-\unab\cdot \uR)
-( -\partial_0 \uoR +\unab\times \uoS )\cdot
( -\partial_0 \uR +  \unab\times\uS)\right)
\]

%The expansion in three-vector calculus involving $\uS, \uR$ and $\uG$ (with scalar $G_0$) and complex conjugates follows from (\ref{eq:NewBetaList}), (\ref{eq:ASTLabels}),
%matrix forms such as $(\ul \cdot \unab ) \Rightarrow -\unab \times $\, , and 
%use of partial integration. 
%It is evident from (\ref{eq:L2FG_DKP}) that 
Here the four-vector components of $\Phi$ have again been eliminated using their equations of motion. The action is proportional to the (complex) Lorentz invariant length, when re-expressed in terms of derivatives of $\uS, \uR$\,. We find 
\begin{align}
\label{eq:L2FG}
\textstyle{\frac 14}{\mathscr L}^{KR} =&\, 
\Phibarbar \ri \widecheck{\beta}{}^\mu \partial_\mu\Phi -m \Phibarbar(1+\beta_5^2)\Phi  = \Gb{}^\mu G_\mu \\
\mbox{where}\quad G_\mu := &\,\textstyle{\frac 12}\varepsilon_{\mu\nu\rho\sigma}\partial^\nu Z^{\rho\sigma}\,\nonumber\\
\mbox{with}
\qquad
G_0=&\, -\unab\cdot \uR 
\,, \quad \uG= -\partial_0 \uR + \unab \times \uS \,.\nonumber
\end{align}
\mbox{}\hfill $\Box$

As pointed out in section \ref{sec:Lorentz_kinetic}\,, under the special Lorentz group, an antisymmetric tensor field $Z_{\mu\nu}$ can be projected into two irreducible, so-called self-dual and anti self-dual components, using the Hodge dual
\[
{}^*Z_{\mu\nu}:= \textstyle{\frac 12} \varepsilon_{\mu\nu\rho\sigma}Z^{\rho\sigma} =\pm \ri Z_{\mu\nu}\,.
\]
Individual terms in local actions are no longer separately fixed under parity, but the contributions of the 
(anti) self-dual components to the total count of invariants can easily be inferred
from equations (\ref{eq:SO31Count})-(\ref{eq:GL4O31Count}) above. For imposing duality restrictions  
on the DKP 10-component wave-function, it is evident that the pseudo-scalar $\beta_5$ and Hodge dual projection
(both of whose eigenvalues are $\pm \ri$) play analogous roles, and indeed 
\begin{align}
\label{eq:SelfDualInDKP}
{}^*Z_{\mu\nu} := \pm \ri Z_{\mu\nu} \quad &\, \Leftrightarrow \beta_5 \Phi^{\wedge} = \pm \ri \Phi^{\wedge} 
\end{align}
where $\Phi^{\wedge} $ is the invariant projection via ${\mathbb P}^{\wedge} :=-\beta_5^2 $ on to the upper
(antisymmetric tensor) components\footnote{With ${\mathbb P}^{\vee} :=1 - {\mathbb P}^{\wedge} = 1+\beta_5^2 $
the lower projection.}  of $\Phi$\,. In terms of the components
in equation (\ref{eq:ASTLabels}) we find correspondingly $\uR = \pm \ri \uS$\,.

Note that, for a 
self-dual or anti self-dual wave function with $\beta_5 (\beta_5^2) \Phi = \pm \mathrm{i}
\beta_5^2 \Phi$, we have\footnote{Note $\beta_5^3 = -\beta_5$ and $\{\eta,\beta_5\}=0$\,.}
\begin{align*}
\Phibar \beta_5 \Phi \!=\! &\,- \Phi^\dagger \eta \beta_5  (\beta_5^2) \Phi \!=\!  
\mp \mathrm{i}\Phi^\dagger \eta (\beta_5^2) \Phi \\
\equiv &\, %+\Phi^\dagger \beta_5\eta \Phi \!=\!  
+\Phi^\dagger (\beta_5)^2\beta_5\eta \Phi \!=\!  \pm \mathrm{i}
\Phi^\dagger\eta  (\beta_5)^2 \Phi \,,
\end{align*}
whence $\Phibar \beta_5 \Phi =0=\Phibar \beta_5^2 \Phi $\,, or equivalently, $\Phibar \Phi \equiv \Phibar (1\!+\! \beta_5^2) \Phi $\,, expressing the fact that only the projection onto the vector part is present, as is evident from the following explicit calculations.

\noindent
\textbf{Case 2: Antisymmetric tensor field \cite{AvdeevChizhov1994,Chizhov2011}:}\mbox{}

\noindent
Take for example a complex anti self-dual DKP wave-function 
(${}^*Z_{\mu\nu}=-\ri Z_{\mu\nu}$\,, 
$\beta_5 \Phi{}^{\wedge}= -\ri \Phi{}^{\wedge}$\,) with
\[
\Phi = {}^\top\big( \uS, -i\uS, +\uG, G_0\big)\,,
\quad 
\Phibar = \Phi^\dagger \eta\, = \big( \uSb, -\ri\uSb, +\uGb, -{\Gb}_0\big)\,.
\] 
Consider the action 
\begin{align}
\label{eq:ASTdef}
\mathscr{L}^{AC} := &\,\Phibar\widecheck{\beta}{}^\mu \textstyle{\frac 12}\overleftrightarrow{\partial_\mu} \Phi  + m 
\Phibar\Phi \,.
\end{align}
We have (see appendix \ref{subsec:DKP_Proca} for details)
\begin{align*}
%\mathscr{L}_{AST} = &\, 
%( \uoS,-\ri \uoS,+\uoG,-\oG_0) \left(\! \begin{array}{c} 
%- \unab\times \uG \\
% - \partial_0 \uG + \unab G_0 \\
%  +  \ri \partial_0 \uS -  \unab\times\uS \\
% \ri \unab\cdot \uS
%\end{array} \!\right)\,+ \mathrm{h.c.} + M \Phibar\Phi  \\
\mathscr{L}^{AC} =&\, \th\Big(\uoS \cdot (- \unab\times \uG) +
(-\ri \uoS)\cdot (- \partial_0 \uG + \unab G_0) +
\uoG\cdot (+  \ri \partial_0 \uS -  \unab\times\uS) -\oG_0 (   \ri \unab\cdot \uS )\\
&\, + \uS \cdot (- \unab\times \uoG) +
(+\ri \uS)\cdot (- \partial_0 \uoG + \unab \oG_0) +
\uG\cdot (-  \ri \partial_0 \uoS -  \unab\times\uoS) -G_0 (   -\ri \unab\cdot \uoS )\Big)\\
&\,+ m(\uoG\cdot \uG - \oG_0 G_0)\\
\Rightarrow &\,
\frac{1}{m}\left((\unab\cdot\uoS)(\unab\cdot\uS)
-(\partial_0\uoS -\ri \unab\times \uoS)\cdot
(\partial_0\uS +\ri \unab\times \uS)\right)
\end{align*}
after elimination. With the re-scaling $\uS \rightarrow \sqrt{m} \uS$ we have
\begin{align}
\label{eq:ACfirstorder}
\mathscr{L}^{AC} =&\,\Phibar\widecheck{\beta}{}^\mu \textstyle{\frac 12}\overleftrightarrow{\partial_\mu} \Phi  + m 
\Phibar\Phi =
\partial^\rho \overline{Z}{}_{\rho \mu} \partial_\sigma Z{}^{\sigma \mu} =  \Gb{}^{ \mu} G_\mu\,, \quad \\
%\nn\\
\mbox{where} \qquad G_\mu := &\,\partial^\rho Z_{\rho \mu}\,,
\quad \nn \\
\mbox{with} \quad G_0= &\, \unab\cdot \uG\,,\quad 
\uG =  \partial_0 \uS + \ri \unab \times \uS\,. \nn
\end{align}
in accord with the (complex) Lorentz invariant length of the gradient of the (anti) self-dual complex antisymmetric tensor field 
\cite{LemesRenanSorella1995II},\cite{LemesRenanSorella1995I}\,. \\
\mbox{}\hfill $\Box$

\noindent
\textbf{Case 3: Antisymmetric tensor field (antisymmetric coupling) 
\cite{Thierry_Mieg_Jarvis_2021,Thierry_Mieg_2021}:}\mbox{}

\noindent
The final invariant coupling identified in section \ref{sec:Lorentz_kinetic} above arises from
a scenario in which there is an internal symmetry (possibly local) conferring an
antisymmetric bilinear invariant. Note that in the standard DKP Proca formalism
a kinetic term such as $\Phi^{\top} \eta \beta^\mu \partial_\mu \Phi$ becomes a total
derivative in view of the symmetry of $\eta\beta^\mu$\,; 
in the complex self-conjugate form $\Phi^{\dagger} \eta \ri \beta^\mu \partial_\mu \Phi$
$\equiv \Phibar \ri \beta^\mu \partial_\mu \Phi$\,, this is of course averted. Appending an additional internal
index $\Phi^\ta$\,, and antisymmetric bilinear invariant $\kappa_{\ta\tb}=-\kappa_{\tb\ta}$ however, again with an anti self-dual
DKP wave-function, permits the alternative action\footnote{
The possible combination $\eta' \widecheck{\beta}{}^\mu$ also has the correct ingredients
for a variant $\Phi^{\top} \eta' \widecheck{\beta}{}^\mu \partial_\mu \Phi$.} in terms
of $\Phi^c:= \Phi^\top \eta$\,,
\begin{align}
\label{eq:CP} 
\mathscr{L}^{CP} := &\,
\Phi^c{}^{\texttt{a}}(\beta^\mu \partial_\mu \Phi^{\texttt{b}}\kappa_{\ta \tb   })\!+\!
\mathrm{h.c.}\!-\! \ri m \Phibar^\ta \Phi^\tb \kappa_{\ta \tb   }\,.
\end{align}
Expanding components in terms of three-vector notation as above, 
%as in equations
%(\ref{eq:L2FG_DKP}), (\ref{eq:L-ASTexpansion}) above, \\
\begin{align*}
\mathscr{L}^{CP} := &\,
2\uG^\ta\cdot (\partial_0 \uS^\tb \!+\!\ri \unab \times \uS^\tb)\kappa_{\ta\tb}
+ 2G^\ta_0(\unab \cdot\uS^\tb)\kappa_{\ta\tb}+
2\uoG^\ta\cdot (\partial_0 \uoS^\tb \!-\!\ri \unab \times \uoS^\tb)\kappa_{\ta\tb}
+ 2\oG^\ta_0(\unab \cdot\uoS^\tb)\kappa_{\ta\tb}\\
&\,+\ri m(\oG^\ta_0 G^\tb_0-\uoG^\ta \cdot \uG^\tb)\kappa_{\ta\tb}\\
\Rightarrow&\,
-\frac{4}{\ri m}(\unab \cdot\uoS^\ta)(\unab \cdot\uS^\tb)\kappa_{\ta\tb}+\frac{4}{\ri m}(\partial_0 \uoS^\ta \!-\!\ri \unab \times \uoS^\ta)
\cdot(\partial_0 \uS^\tb \!+\!\ri \unab \times \uS^\tb)\kappa_{\ta\tb} 
\,, 
\end{align*}
again after elimination\,. With the field redefinition 
$\uS \rightarrow \th \ri\sqrt{m}\uS$\,, we have finally 
\begin{align}
\mathscr{L}^{CP} := &\, 
\Phi^c{}^{\texttt{a}}(\beta^\mu \partial_\mu \Phi^{\texttt{b}}\kappa_{\ta \tb   })\!+\!
\mathrm{h.c.}\!-\! \ri m \Phibar^\ta \Phi^\tb \kappa_{\ta \tb   }\,=\ri\partial^\rho \overline{Z}{}^\ta_{\rho \mu} \partial_\sigma Z{}^{\tb\,\sigma \mu} \kappa_{\ta\tb}\\
= &\, \ri\Gb{}^{\ta\mu} G^b_\mu\kappa_{\ta\tb}\,, 
\quad
\mbox{where} \quad
\quad G^\ta_\mu :=\partial^\rho Z^\ta_{\rho\mu}\,,\quad \mbox{with} \nn \\
G_0^\ta = &\,  \unab\cdot \uS \,, \quad \uG^\ta = \partial_0 \uS^\ta +\ri
\unab \times \uS^\ta \,.\nn
\end{align}
\mbox{}\hfill $\Box$

\noindent
\section{Conformal invariance.}
\label{sec:ConfDKPproof}
To complete the discussion of symmetry aspects of the physical antisymmetric tensor models, reviewed in 
section \ref{sec:Lagrangian_section} above, we here provide a demonstration of the conformal invariance both of 
the antisymmetrically coupled pseudoscalar ${\mathscr L}^{CP}$\, (equation
 (\ref{eq:CP_complex})) here using its first order four dimensional DKP formulation 
(table (\ref{tab:DKP_2FGASSTlist}), equation (\ref{eq:CP})  )\,. A similar calculation obtains for the symmetrically coupled scalar ${\mathscr L}^{AC}$\, 
( table (\ref{tab:DKP_2FGASSTlist}) and equations (\ref{eq:AC_complex})\,, (\ref{eq:ACfirstorder})  )\,.

We adapt to the present case the analysis of Jackiw and Pi \cite{Jackiw_2011}\,, which gave a convenient criterion for verifying conformal invariance in a scale-invariant system. In the first order formulations however it is 
 necessary to adjust for the fact that
the scaling behaviour of the 10 component DKP wave function (\ref{tab:candidates})\, is no longer a diagonal multiplier, but must be taken as
$\ttD :=\ttd \mathrm{diag}(1,1,2,2)$ in $3\!+\!3\!+\!3\!+\!1$ block form\footnote{We take the parameter $\ttd \rightarrow 1$ henceforth. By contrast, in the traditional Kemmer version of the Proca massive vector field we would have $\ttD' :=\mathrm{diag}(2,2,1,1)$\,.}.
Given that $\beta_5^2 =\mathrm{diag}(-1,-1,0,0)$, other
Lorentz invariants can be written similarly as 
\[
\quad \beta_\mu\beta^\mu =
\mathrm{diag}(2,2,3,3)= (3\!+\!\beta_5^2)\,, 
\quad \ttD = (2\!+\!\beta_5^2)\,,
\]
with the projectors ${\sf Id} = (1+\beta_5)^2 + (-\beta_5)^2$\,. 

Checking  the scale invariance of the first order Lagrangian (\ref{eq:CP})\,, we have\footnote{Recall that $\Phibar\Phi = \Phibar(1\!+\!\beta_5^2)\Phi$ for (anti) self-dual wave functions, $\beta_5^4 = -\beta_5^2$\,, and that $\{\beta_5^2,\beta^\sigma\} =-\beta^\sigma$\,.}
\begin{align*}
\delta\big(\Phibar(1\!+\!\beta_5^2)\Phi\big)=&\,
= \Phibar\{\ttD,(1\!+\!\beta_5^2)\}\Phi =\Phibar\{(2\!+\!\beta_5^2),(1\!+\!\beta_5^2)\}\Phi=4(\Phibar(1\!+\!\beta_5^2)\Phi)\,,\\
\delta(\Phi^c \beta^\mu \partial_\mu \Phi) =&\, 
(\Phi^c \beta^\mu \partial_\mu \Phi) + 
(\Phi^c \{\beta^\mu, (2\!+\!\beta_5^2)\} \partial_\mu \Phi)\\
=&\,
5(\Phi^c \beta^\mu \partial_\mu \Phi) +(\Phi^c \{\beta^\mu, \beta_5^2\} \partial_\mu \Phi) = 4(\Phi^c \beta^\mu \partial_\mu \Phi)\,.
\end{align*}
\noindent
The criterion of Jackiw and Pi \cite{Jackiw_2011} for conformal  invariance is that  the field virial\,,
\begin{align*}
J^\sigma = &\,
\frac{\delta {\mathscr L}}{\delta \partial^\mu \varphi}
\big(\ttd \eta^{\mu\sigma} - \Sigma^{\mu \sigma} \big)\varphi\,,
\end{align*}
is a divergence, $J^\sigma = \partial_\mu W^{\mu\sigma}$\,. Here the sum is over all fields $\varphi$ of scaling dimension $\ttd$\,, and $\Sigma^{\mu \sigma} $ are the Lorentz group generators applied to the field multiplet $\varphi$\,.

We assume here that the method of \cite{Jackiw_2011} simply goes through, again replacing the scaling dimension by the operator $\ttD$, and also with
use of the reducible representation of the Lorentz group generators, provided by the Kemmer $10\times 10$ matrices. Proceeding with the functional differentiation of (\ref{eq:CP})\,
we have
\begin{align*}
J^\sigma = &\,\Phi^c\beta_\mu \big(\ttD \eta^{\mu\sigma} + {[}\beta^\mu,\beta^\sigma{]}\big)\Phi\\
\mbox{where}\qquad
\beta_\mu \big(\ttD \eta^{\mu\sigma} + {[}\beta^\mu,\beta^\sigma{]}\big)
=&\, \beta^\sigma (2+\beta_5^2) +  \beta_\mu\beta^\mu\beta^\sigma
- \beta_\mu \beta^\sigma\beta^\mu\\
=&\, \beta^\sigma (2+\beta_5^2)  +(3+\beta_5^2) \beta^\sigma -2 \beta^\sigma
\equiv 3\beta^\sigma + \{\beta_5^2,\beta^\sigma\} = 2 \beta^\sigma
\end{align*}
where we have assumed $\ttd = 1$ and used $\{\beta_5^2,\beta^\sigma\} =-\beta^\sigma$\,. Hence in antisymmetric coupling
\[
J^\sigma =2 \kappa_{\tta\ttb}(\Phi^\tta)^\top \eta \beta^\sigma \Phi^\ttb + h.c. \equiv 0+0
\]
because $( \eta \beta^\sigma)$ is symmetric and $\kappa_{\tta\ttb}$ is antisymmetric\,.

A similar computation applies to the (symmetric) Avdeev-Chizhov Lagrangian 
(\ref{eq:ACfirstorder}) for scaling invariance, as 
$\{\beta_5^2,\widecheck{\beta}{}^\sigma\} =-\widecheck{\beta}{}^\sigma$\,. A detailed expansion shows that the field virial is
now proportional to 
%$\kappa_{\tta\ttb}\Phibar{}^\tta \widecheck{\beta}{}^\sigma \Phi^\ttb + h.c.$\,, 
$\Phibar{} \widecheck{\beta}{}^\sigma \Phi  + h.c.$\,, 
which again vanishes, but now due to the {\emph{antisymmetry}} 
of $(\eta \widecheck{\beta}{}^\sigma)$\,. %\hfill $\Box$

\end{appendix}
\vfill
\end{document}